\documentclass[a4paper,10pt]{article}

\usepackage{jheppub} 

\usepackage[T1]{fontenc} 

\usepackage{bm}
\usepackage{latexsym}
\usepackage{dcolumn}
\usepackage{amsmath,amsfonts,amssymb, amsthm}
\usepackage{graphicx,epsfig}
\usepackage{environ}
\usepackage{mathtools}
\usepackage{braket}
\usepackage{float}
\usepackage{mathrsfs}
\usepackage{esint}
\usepackage{amsmath}
\usepackage{tensor}
\usepackage{fancyhdr}
\usepackage{subcaption}
\usepackage{hyperref}
\usepackage{graphicx,epstopdf}
\usepackage{capt-of}
\usepackage{tikz}
\usepackage{scalerel}
\usepackage{fixmath}
\usepackage[bb=boondox]{mathalpha}

\hypersetup{
	colorlinks   = true, 
	urlcolor     = blue, 
	linkcolor    = blue, 
	citecolor   = red 
}

\def\a{\alpha}
\def\b{\beta}
\def\g{\gamma}
\def\d{\delta}

\def\p{\partial}

\def\T{\Theta}

\def\k{\kappa}

\def\be{\begin{equation}}
\def\ee{\end{equation}}
\def\ba{\begin{align}}
\def\ea{\end{align}}

\def\G{\Gamma}

\newcommand{\levicivita}{}
\def\levicivita#1#{\tensor#1{\epsilon}}

\def\sptm{(\mathcal{M}, \textbf{g}, \hat{\bm{\nabla}})}
\def\hech{\mathcal{H}}

\def\hcvd{\hat{\nabla}}
\def\cvd{\nabla}

\def\thht{\hat{\Theta}}
\def\ti{\mathbb{T}}

\def\hw{\hat{\omega}}
\def\hhw{\hat{\Omega}}

\def\hchi{\hat{\chi}}
\def\lil{\pounds_{\bm{l}}}
\def\lik{\pounds_{\bm{k}}}
\def\thdl{\overset{(d)}{\hat{\theta_{\bm{l}}}}}
\def\thdk{\overset{(d)}{\hat{\theta_{\bm{k}}}}}
\def\shdf{{^{(\bm{l},d)}{\sigma}}}
\def\shdl{{^{(\bm{l},d)}{\sigma_{ab}}}}

\def\shdk{{^{(\bm{k},d)}{\sigma_{ab}}}}

\def\xih{\hat{\Xi}}
\def\fr{\frac{1}{2}}
\def\hb{\hat{\mathcal{B}}}

\def\shbl{{^{(\bm{l},B)}{\sigma_{ab}}}}

\def\whbl{{^{(\bm{l},B)}{\omega_{ab}}}}

\def\hr{\hat{R}}

\def\hsvd{\hat{\mathcal{D}}}

\def\hr{\hat{R}}

\def\hp{\hat{\mathscr{P}}}
\def\mt{\overset{(m)}{T}}

\def\mt{T^{(m)}}
\def\hg{\hat{G}}
\def\hgm{\hat{\Gamma}}

\def\aem{\mathcal{A}_{\text{m}}}

\def\m4{\mathcal{O}(\epsilon^4)}

\def\hphi{\hat{\Phi}}
\def\kt{\bm{k} \cdot  \mathbbb{T}}
\def\ktb{k_b \ti^b}
\def\stth{\overset{*}{\theta}}
\def\stsig{\overset{*}{\sigma}}
\def\stome{\overset{*}{\omega}}
\def\tit{\tilde{t}}
\def\qq{q^i_{~m}q^j_{~n}}
\def\qtl{q^{rs}T_{rts}l^t}
\def\spdt{\hat{\mathcal{D}}_{\bar{t}}}
\def\tmu{\tilde{\mu}}
\def\tnu{\tilde{\nu}}
\def\tal{\tilde{\alpha}}
\def\tbe{\tilde{\beta}}
\def\oeq{\overset{\hech}{=}}

\title{Possible fluid interpretation and tidal force equation on a generic null hypersurface in Einstein-Cartan theory}
\author[a]{Sumit Dey, Bibhas Ranjan Majhi}
 \affiliation[a]{Department of Physics, Indian Institute of Technology Guwahati, Guwahati 781039, Assam, India.}
\emailAdd{dey18@iitg.ac.in}
\emailAdd{bibhas.majhi@iitg.ac.in}
\abstract{The dynamical evolution of the Hajicek $1$-form is derived in Einstein-Cartan (EC) theory. We find that like Einstein theory of gravity, the evolution equation is related to a projected part of the Einstein tensor $(\hg_{ab})$ on a generic null surface $\hech$, particularly $\hg_{ab}l^a q^b_{~c}$, where $l^a$ and $q^a_{~c}$ are the outgoing null generators of $\hech$ and the induced metric to a transverse spatial cross-section of $\hech$ respectively. Under the {\it geodesic constraint} a possible fluid interpretation of this evolution equation is then proposed. We find that it has the structure which is reminiscent to the {\it Cosserat generalization} of the Navier-Stokes fluid provided we express the dynamical evolution equation of the Hajicek $1$-form in a set of coordinates adapted to $\hech$ and in a local inertial frame. An analogous viewpoint can also be built under the motive that the usual material derivative for fluids should be replaced by the Lie derivative. Finally, the tidal force equation in EC theory on the null surface is also derived.}

\date{\today}

\begin{document}
\maketitle
\flushbottom
\noindent
\section{Introduction}
	Over the past few decades there have been mounting theoretical evidence that the field equations of gravity could be \textit{emergent} in nature, much like that of elasticity theory, fluid mechanics or gas dynamics \cite{Padmanabhan:2003gd, Padmanabhan:2007en, Padmanabhan:2009vy, Padmanabhan:2009kr, Padmanabhan:2010xe, Padmanabhan:2010xh, padmanabhan2010gravitation,Kolekar:2011gw, Kolekar:2011bb, Padmanabhan:2013xyr, Padmanabhan:2013nxa, Chakraborty:2014rga, Parattu:2013gwa, Chakraborty:2014joa, Padmanabhan:2014jta}. The macroscopic gravitational dynamics can be viewed as a result of coarse-graining leading to the the thermodynamic limit of the underlying statistical mechanics of the ``atoms'' of spacetime. Analogously, the dynamics of gravity can also be considered as the long wavelength hydrodynamic limit of the underlying ``fluid/elastic-model'' system. The phenomenon of emergence of gravity is used in the specific context of the respective gravitational field equations rather than that of emergence of the geometry of space and time itself. For example, in the gauge/gravity theory, it has been argued that classical spacetimes emerge as a consequence of quantum entanglement \cite{VanRaamsdonk:2010pw}.  The indication of such an emergent nature (in the purview of gravitational field equations) have been supported in the literature through various directions by the works of several authors (originating from the work of Sakharov \cite{sakharov1967vacuum}).

	In this context, the emergence of gravitational dynamics via the field equations from underlying thermodynamical relations or fluid equations have been developed and explored. It was shown by Jacobson \cite{Jacobson:1995ab}, that the Einstein field equations can be obtained from an underlying local constitutive relation (specifically the Clausius identity $\d S = \frac{\d Q}{T}$) applied to local causal Rindler horizons ``\textit{in equilibrium}''. The temperature of the horizon is assumed to be the Unruh temperature with the entropy variation being proportional to area change of the cross-sectional area of the horizon. Jacobson's formalism has also been extended for local causal horizons in the non-equilibrium case \cite{Eling:2006aw, Chirco:2009dc, Dey:2017fld} and for other modified theories of gravity \cite{Chirco:2009dc, Dey:2017fld}.
		
		 Padmanabhan {\it et-al} \cite{Padmanabhan:2002sha, Padmanabhan:2003gd, Padmanabhan:2009vy, Kothawala:2007em, Paranjape:2006ca,Chakraborty:2015aja} have interpreted in a consistent fashion, the gravitational field equations with respect to (w.r.t) a generic null hypersurface for a wide class of gravity theories as thermodynamic relations. The action functional of gravity lends itself a thermodynamic interpretation \cite{Padmanabhan:2002sha, Padmanabhan:2004fq, Mukhopadhyay:2006vu, Kolekar:2010dm} and the field equations can be obtained from the extremization of a thermodynamic extremum principle. Padmanabhan has shown \cite{Padmanabhan:2007en, Padmanabhan:2007xy} that the Einstein and the Lanczos-Lovelock field equations can be obtained via the extremization of an entropy functional constructed from the null vectors in the spacetime.

		In a relatively old work, Damour \cite{Damour:1979wya, damour1982surface} showed that the Einstein field equations have the same status as that of the equations of fluid mechanics in the context of a black hole event horizon. Projecting the Einstein field equations onto the horizon, Damour showed that it took a form similar (however not exactly) to the Navier-Stokes (NS) equation. This allowed the interpretation of the black hole horizon as a dissipative membrane which later gave way to the Membrane paradigm \cite{ thorne1986black}. The result was generalized to the case of a generic null hypersurface in Einstein gravity. This led to the Damour-Navier-Stokes (DNS) equation \cite{Gourgoulhon:2005ng, Padmanabhan:2010rp} for a viscous null fluid. It was shown by Padmanabhan \cite{Padmanabhan:2010rp}, that when the DNS equation is viewed in a boosted inertial frame, then it reduces exactly to the NS equation. The identified fluid variables for the DNS equation are related to the geometric/kinematical quantities of the null hypersurface constructed in the spacetime. This allows the conceptualization that a general null hypersurface behaves as a dissipative null fluid in Einstein gravity.	
		
		These thermodynamical/physical equivalences or connections would possibly not have arisen had the gravitational field equations not emerged from an underlying micro-structure. Now, if gravity is indeed emergent, then such vivid equivalences/correspondences should transcend to other other theories of gravity. In this paper, we will focus exclusively on the Einstein-Cartan (EC) theory \cite{ASENS_1923_3_40__325_0, Hehl:1976kj}. The Riemann-Cartan (RC) spacetime is the geometric backdrop for the EC theory. The EC theory includes an additional intrinsic spin of particle(s) in the geometrization of spacetime. This causes a non-zero torsion in the spacetime geometry leading to the connection being not explicitly symmetric. The geometry or the metric is sourced by the energy-momentum tensor whereas the spin degrees of freedom or torsion is sourced by the spin-angular momentum tensor. The relevant gravitational field equations in the EC theory is the Einstein-Cartan-Kibble-Sciama (ECKS) equations \cite{kibble1961lorentz, sciama1962, sciama1964physical, sciama1964erratum, hehl1971does, de1986introduction, de1994spin, kleinert1989gauge, Poplawski:2009fb, Shapiro:2001rz, Obukhov:2006gea}. Our quest will be hinged on the fact as to whether the ECKS equations have any thermodynamic or fluid interpretation w.r.t a general null hypersurface constructed in the RC spacetime. To that extent, the following projections of the tensor $\hg_{ab}$ (analog of the Einstein tensor $G_{ab}$ in the RC spacetime) on the null hypersurface would be very crucial.
		 \begin{enumerate}
		 	\item $\hg_{ab}l^a l^b$ (where $l^a$ are the null generators of the null surface) is related to the evolution of outgoing expansion scalar and hence leads to the null Raychaudhuri equation (NRE) \cite{Poisson:2009pwt,Gourgoulhon:2005ng}. The NRE was a crucial input in Jacobson's analysis to obtain the Einstein field equations from the Clausius identity \cite{Jacobson:1995ab}. The Raychaudhuri equation as well as the dynamical evolution equations for the shear and vorticity for both timelike and null curves established for the RC spacetime has already been explicitly derived in \cite{PhysRevD.104.084073, PhysRevD.96.024021, PhysRevD.58.044021, Dey:2017fld, Dey:2022qqp}. The corresponding emergence of the ECKS field equations from a local constitutive thermodynamical relation established for  causal Rindler horizons has been shown in \cite{Dey:2017fld} and hence will not be pursued here.
		 	\item $\hg_{ab}k^a l^b$ (where $k^a$ is the auxiliary null vector field to the null surface) is related to the dynamical evolution of the ingoing expansion scalar. The quantity $G_{ab}k^a l^b$ was a crucial input to show (in a covariant fashion) that the Einstein field equations expressed w.r.t a generic null surface assumes a thermodynamic identity very similar to the first law of thermodynamics \cite{Dey:2020tkj} \footnote{it is the covariant formulation of earlier investigations \cite{Padmanabhan:2002sha, Padmanabhan:2003gd, Padmanabhan:2009vy, Kothawala:2007em, Paranjape:2006ca,Chakraborty:2015aja}. In this connection the idea of assigning temperature on a generic null surface has been put-forwarded from the physical point of view in \cite{Dalui:2021sme}.}. Extending to the case of EC theory, the authors of this paper showed that such thermodynamical interpretation can indeed be alluded to the the ECKS equations \cite{ Dey:2022qqp}. Therefore, we will not pursue this issue here. The thermodynamic interpretation to the gravitational theory via $G_{ab}k^a l^b$ was also applied to Scalar-Tensor theory in \cite{Dey:2021rke}.
		 	\item $\hg_{ab}l^a q^b_{~c}$ (where $q_{ab}$ is the induced metric on a transverse spacelike cross section of the null surface) is related to the dynamical evolution of the Hajicek $1$-form of the null surface \cite{Gourgoulhon:2005ng}. For the case of Einstein gravity, $G_{ab}l^a q^b_{~c}$ leads to the Lie evolution of the Hajicek $1$-form along the null generators (of the null surface) which is then interpreted as the DNS equation \cite{Gourgoulhon:2005ng, Padmanabhan:2010rp}. It also works in scalar-tensor theory of gravity as well \cite{Bhattacharya:2020wdl}. In this paper, we will focus exclusively whether $\hg_{ab}l^a q^b_{~c}$ would enable us to attribute to the ECKS field equations any fluid/elastic continuum model interpretation. In particular, under the geodesic constraint we will be able to see when we write down the dynamics of the Hajicek $1$-form in a coordinate system adapted to the null surface $\hech$, the structure we get is quite similar to the Cosserat generalization of the NS fluid (under appropriate identifications of the fluid variables with the kinematical variables of $\hech$ and the external force density). In fact, we will see that w.r.t a local inertial frame, the above dynamical equation will indeed reduce to the Cosserat fluid equation. The Cosserat fluid is a real world fluid dynamical system that incorporates intrinsic angular momentum.
		 	
		 	Even though, much of our analysis is based on providing a possible fluid interpretation for the ECKS field equations, we also present the tidal force equation for the null surface $\hech$ in its most generic sense. The only imposition, that we have made while arriving at the tidal equation is that the geodesic null congruence generates an integrable hypersurface, i.e it satisfies the Frobenius identity. In doing this analysis, we are also led to derive to the NRE for the given integrable hypersurface $\hech$ in the RC spacetime. The tidal equation and the NRE analyzed for the EC theory furnish a part of the optical scalar equations obtained under the Newman-Penrose formalism.

		 	The organization of the paper is as follows. In the next section, we provide a very brief review of the RC spacetime, the construction of a generic null hypersurface $\hech$ in it and the relevant kinematics of $\hech$. In section \ref{summary2}, we spell out only the mere essentials of the EC theory and its field equations that would be required for our analysis. In section \ref{dynamics1}, we begin our in-depth analysis of the dynamical evolution of the Hajicek $1$-form and show that its dynamics is indeed governed by the projection component $\hg_{ab}l^a q^b_{~c}$. In the next section \ref{interpretation}, we try to argue whether the ECKS field equation w.r.t $\hech$ via the component $\hg_{ab}l^a q^b_{~c}$ can be attributed any possible fluid dynamical/elastic theory interpretation. Here, we come to the conclusion that the resulting dynamics cannot be compared with the DNS fluid. The dynamics has rather an analogy with the Cosserat generalization of the NS fluid which we try to make more precise in a boosted local inertial frame.  We do this to argue whether the dynamics of the ``null fluid'' established on $\hech$ has connections/analogy with some real world fluid scenarios. For the sake of completeness, we also present in section \ref{tidalx} the tidal force equation governing a congruence of null geodesics in the RC spacetime. We then conclude in \ref{conclusion} and in the appendices provide detailed derivations of some of the expressions used in the text.
		 	
		 	A note about our notations, dimensions and conventions. We adopt the metric signature $(-,+,+,+)$ and work in $d= 4$ spacetime dimensions. We will use the geometrized unit system where $c$, $\hbar$ and $G$ have been set to one. The bulk spacetime indices are designated by the lowercase Latin alphabets $a,b, \cdot \cdot \cdot$. The spatial coordinates on a time constant slice are designated by the Greek alphabets $\mu ,\nu,\cdot \cdot \cdot$. The coordinates on the null hypersurface are designated by Greek alphabets, with a tilde on them, $\tmu, \tnu, \cdot \cdot \cdot$. The spatial coordinates on the two dimensional spatial cross section $S_t$ of $\hech$ are designated by the uppercase Latin alphabets $A, B, \cdot \cdot \cdot$. All kinematical and dynamical quantities associated with $\hech$ in the RC spacetime will be designated with a hat on them. The equivalent quantities in the usual spacetime without torsion will be unhatted.
		 	
		 \end{enumerate}

\section{Summary of kinematics of an integrable null hypersurface in the RC spacetime: a short review}\label{summary1}
In this section, we give a very brief overview of the relevant kinematical quantities pertaining to the general construction of an integrable null hypersurface in the RC spacetime. The detailed description of the geometry of a null hypersurface in RC spacetime has been carried out in \cite{Gourgoulhon:2005ng} and hence here, we will be content in exposing only the salient features. These will be needed to achieve the present goals.

Our ambient spacetime $\sptm$ is the RC spacetime provided with a general metric compatible affine connection i.e $\hcvd_a g_{bc} = 0$. In the coordinate basis, the torsion tensor is defined as,
\begin{align}
T^a_{~bc} \equiv \hgm^a_{~bc} - \hgm^a_{~cb}~,
\label{torsion}
\end{align}
The contorsion tensor $K^a_{~bc}$ then follows as,
\begin{align}
\hgm^a_{~bc} = \G^a_{~bc} + K^a_{~bc} = \G^a_{~bc} +\fr \Big(T^a_{~bc} + T_{b~~c}^{~a} + T_{c~~b}^{~a}\Big)~.
\label{contorsion}
\end{align}
Another relevant quantity of interest is the modified torsion tensor $S^a_{~bc}$  :
\begin{align}
S^a_{~bc}  \equiv T^a_{~bc} + \d^a_{~b}T_c - \d^a_{~c}T_b ~,\label{modifiedtorsion}
\end{align}
where $T_b \equiv g^{ac}T_{abc}$ is the contraction of the first and third index of the torsion tensor.
In our chosen convention, the Riemann tensor of $\sptm$ is defined w.r.t the affine connection $\hgm^a_{~bc}$ as,
\begin{align}
\hr^a_{~bcd} \equiv \p_c \hgm^a_{~db} - \p_d \hgm^a_{~cb} + \hgm^a_{~ci} \hgm^i_{~db} - \hgm^a_{~di}\hgm^i_{~cb} ~.
\end{align}
Just like the case of the spacetime ($\mathcal{M}, \bm{g}, \bm{\nabla}$) provided with the metric compatible symmetric Levi-Civita connection ($\bm{\nabla}$), we define the analog of the Einstein tensor in the RC spacetime as,
\begin{align}
\hg_{ab} = \hr_{ab} - \fr g_{ab} \hr ~.
\end{align} 
Due to the presence of torsion, the tensor $\hg_{ab}$ fails to be divergenceless,
\begin{align}
\hcvd_a \hg^a_{~b} = - T^k_{~ab} \hr^a_{~k} + \fr T^{kad} \hr_{adkb} ~.
\end{align}
Now, we consider the decomposition of curvature tensor $\hr_{abcd}$ in the RC spacetime into a Riemannian part $(R_{abcd})$ and an extra part that involves the torsion contributions. It can be shown that,
\begin{align}
\hr^a_{~bcd} = R^a_{~bcd} + (\hcvd_c K^a_{~db} - \hcvd_d K^a_{~cb}) + T^i_{~cd}K^a_{~ib} + (K^i_{~cb}K^a_{~di} - K^i_{~db}K^a_{~ci}) ~.
\label{curv1}
\end{align}
Similarly, taking the contraction to yield the Ricci tensor, it can be easily verified that,
\begin{align}
\hr_{ab} = R_{ab} + \hcvd_i K^i_{~ba} + \hcvd_b T_a + T^i_{~jb} K^j_{~ia} + K^i_{~ja} K^j_{~bi} + T_i K^i_{~ba} ~.
\label{curv2}
\end{align}

We will now consider the construction of a general null hypersurface $\hech$ in this RC spacetime. The null hypersurface $\hech$ is defined via the scalar field $u(x^a) = 0$, with the null normal $l_a$ to $\hech$ being,
\begin{equation}
l_a = - e^{\rho} \partial_a u = -e^{\rho} \hat{\nabla}_a u ~,
\label{nullnormal}
\end{equation} 
where $\rho$ is a smooth scalar field defined on $\hech$. Following Carter \cite{carter1997extended}, we will consider not just a single null hypersurface, but rather a family of such null hypersurfaces $u(x^a) = \text{constant}$. Out of this family, our chosen $\hech$, defined by $u(x^a) = 0$ is just a particular member. So we have foliated our spacetime $\sptm$ in an open neighborhood of $\hech$ by a family of null hypersurfaces. The details of such a construction have been explained in \cite{Gourgoulhon:2005ng, Dey:2022qqp} and will not be pursued here. The null generators $\bm{l}$ of $\hech$ satisfy the following relation,
\begin{equation}
l^a \hcvd_a l_b - T_{abc}~l^a l^c = l^a \cvd_a l_b = \kappa l_b ~,
\label{parrl1}
\end{equation}
where $\kappa$ is the non-affinity parameter associated with the null geodesics $\bm{l}$ and $\bm{\nabla}$ is the standard Levi-Civita connection. We observe that the null generators $\bm{l}$ are not autoparallel (w.r.t $\bm{\hcvd}$), even though they are geodesics (w.r.t $\bm{\nabla}$). Only when the condition $\ti_b \equiv T_{abc}l^a l^b = 0$ is satisfied, do we have $l^a \hcvd_a l^b = \kappa l^b$. The restriction $\ti_a = 0$ on the torsion tensor for null generators $\bm{l}$ in the RC spacetime is what we call the \textit{geodesic constraint}. We would also like to consider a $3+1$ foliation of the null family $u(x^a) = \text{constant}$ by a stack of spacelike hypersurfaces $\Sigma_t$ defined as $t(x^a) = k (\text{constant})$. We will assume that these spacelike hypersurfaces do not intersect with each other. Hence, an open neighborhood of $\sptm$ in the vicinity of $\hech$ can be parametrized by well defined local set of coordinates $x^a  = (t, x^{\mu})$, where $x^{\mu} = (x^1,x^2, x^3)$ are the spatial coordinates installed on the spacelike surface $\Sigma_t$. The coordinate time evolution vector $\bm{t} = \bm{\p_{t}}$ that connects the same spatial points $y^{\mu}$ on the neighboring spacelike slices satisfy $t^a \p_a t = 1$. The intersection of the family $\Sigma_t$ with our null hypersurface $\hech$ is the stack of transverse $2$-dimensional cross sections $S_t$ defined as $S_t \equiv \hech \cap \Sigma_t $. We also have a unique transverse auxiliary null vector field $\bm{k}$ defined as, 
\begin{equation}
\bm{l} \cdot \bm{k} = -1, ~~~ \bm{k} \cdot \bm{k} = 0 ~~ \text{and} ~~ \bm{k} \cdot \bm{e_{{A}}} = 0 ~,
\label{defnauxiliaryk}
\end{equation}
where $\{\bm{e_{{A}}}\}$ denotes the set of basis vectors established on the tangent space of $S_t$. It can be seen \cite{Gourgoulhon:2005ng}, that the null generators $\bm{l}$ represents a notion of outgoing null vector field w.r.t $\hech$. On the other hand, the auxiliary null field $\bm{k}$ denotes a notion of ingoing vector field w.r.t $\hech$. The orthogonal projection tensor onto the spacelike submanifold $S_t$ is defined as,
\begin{equation}
q_{ab} = g_{ab} + l_a k_b + k_a l_b ~.
\label{projectiontensorSt1}
\end{equation} 

Now, we come to the relevant kinematical description of $\hech$. All such quantities have been developed in details in \cite{Dey:2022qqp}. The first such quantity that we are interested in is the extended second fundamental form $\thht_{ab}$ which in essence quantifies the extrinsic curvature of our null hypersurface. This is defined as,
\begin{equation}
\thht_{ab} = \Big(q^c_{~a} q^d_{~b} \hcvd_d l_c\Big) - \Big(q^c_{~a} k_b \ti_c\Big) ~.
\label{extsecondfundform}
\end{equation}
Notice that due to the presence of nontrivial torsion in the spacetime, the extended second fundamental form is not symmetric \cite{Dey:2022qqp}, i.e.
\begin{align}
\thht_{ba}- \thht_{ab} = q^c_{~a} q^d_{~b} T_{fdc}~l^f + (k_b q^c_{~a} - k_a q^c_{~b}) \ti_c ~.
\label{diffinTheta}
\end{align}  
It is also quite easy to verify that,
\begin{equation}
\thht_{ab} l^a = 0, ~~ \thht_{ab}l^b = q^c_{~a} \ti_c , ~~ \thht_{ab} k^a = 0 ~~ \text{and} ~~ \thht_{ab} k^b = 0 ~.
\label{extfundformprops}
\end{equation} 
The above relation clarifies the fact the second fundamental form is orthogonal to both $\bm{l}$ and $\bm{k}$ only when the geodesic constraint is satisfied. It is only then that the extended second fundamental form becomes a completely spatial bilinear. 

Next, we come to the description of the rotation and the Hajicek $1$-form. The rotation $1$-form $\underline{\bm{\hw}}$ is related to the Weingarten map for the null surface. As a very useful working definition of the rotation $1$-form, it can established \cite{Dey:2022qqp} that,
\begin{equation}
\hw_a = (l^b \hcvd_b k_a) + \hp_a ~,
\label{rotn1forndefn2}
\end{equation}
where we have $\hp_a \equiv T_{bcd} k^b l^c q^d_{~a}$.
Another equivalent definition is,
\begin{equation}
\hw_a = -(k_b \hcvd_a l^b) - l_a k_b (\bm{\hcvd_{k}}l^b) ~.
\label{omega1defn}
\end{equation}
Using the above definitions, it can quite easily be verified that,
\begin{equation}
\hw_a k^a = 0 ~~ \text{and} ~~ \hw_a l^a = \kappa - k_a \ti^a ~.
\label{rotn1formprops}
\end{equation}
The projection of the rotation $1$-form onto the transverse spacelike submanifold $(S_t, \bm{q})$ is defined to be the Hajicek $1$-form $\bm{\underline{\hhw}}$,
\begin{equation}
\hhw_a \equiv q^b_{~a} \hw_b ~.
\label{Omegadefn1}
\end{equation}
The following relation between the rotation and the Hajicek $1$-form can then be established,
\begin{equation}
\hw_a = \hhw_a - \kappa k_a + (k_b \ti^b)k_a ~.
\label{rotnhajireln}
\end{equation}
A quantity of significant interest while dealing with the dynamics of our null hypersurface $\hech$ (to be discussed in the next section) is the spacetime covariant derivative of the null normal $l_b$. The fact that we have foliated an open neighborhood of the spacetime in the vicinity of $\hech$, by a family of null hypersurfaces, allows us to extend the support of $\bm{l}$ to not just only on $\hech$, but rather to locally well defined open neighborhood of spacetime $\sptm$ about the hypersurface $\hech$. Thus the operation $\hcvd_a l_b$ can be properly taken. It can be shown \cite{Dey:2022qqp} that,
\begin{align}
\hcvd_a l_b = \thht_{ba} + \hw_a l_b - l_a (\bm{\hcvd_{k}}l_b ) ~.
\label{nablaalbexpansion}
\end{align}

The next kinematical quantity that comes into our discussion is the deformation rate tensor $\hchi_{ab}$, which quantifies the projection (onto $S_t$) of the rate of change of the metric $q_{ab}$ as its evolves along $\bm{l}$ of $\hech$,
\begin{equation}
\hchi_{ab} = \frac{1}{2}  q^i_{~a} q^j_{~b}\pounds_{\bm{l}} q_{ij} ~.
\label{deformationrate}
\end{equation}
As it can easily be observed from \eqref{deformationrate}, the deformation rate tensor is by definition symmetric and completely orthogonal to $\bm{l}$ and $\bm{k}$. The irreducible decomposition of the spatial bilinear $\hchi_{ab}$ follows as,
\begin{equation}
\hchi_{ab} = \frac{1}{2} q_{ab} ~\overset{(d)}{\hat{\theta_{\bm{l}}}} + {^{(\bm{l},d)}{\sigma_{ab}}} ~,
\label{irredecompdeform}
\end{equation}
where $\thdl$ represents the outgoing expansion scalar corresponding to the null generators $\bm{l}$ of $\hech$ and $\shdl$ is the traceless symmetric shear tensor corresponding to $\hchi_{ab}$. The trace $\thdl$ of the deformation rate tensor is called the outgoing expansion scalar because it correctly quantifies the fractional rate of change of the area measure $\sqrt{q}$ of $S_t$ along $\bm{l}$ \cite{Dey:2022qqp}. It can also be shown that,
\begin{align}
\thdl = \hcvd_a l^a + T_a l^a - \kappa = \nabla_a l^a - \kappa ~.
\label{expansionscalervalue}
\end{align}
The relation between the extended second fundamental form and the deformation rate tensor is as follows \cite{Dey:2022qqp},
\begin{equation}
\hchi_{ab} = \thht_{ba} + k_a q^c_{~b} \ti_c + q^c_{~a} q^d_{~b} K_{fcd}l^f ~.
\label{relnchintheta}
\end{equation} 
Finally, we would like to express the spacetime covariant derivative of the null normal $l_b$ in terms of the deformation rate tensor. To that, it can be established via \eqref{relnchintheta} and \eqref{nablaalbexpansion}, that,
\begin{equation}
\hcvd_a l_b = \hchi_{ab} + \hw_a l_b - l_a (k^i \hcvd_i l_b) - k_a q^c_{~b} \ti_c - q^c_{~a} q^d_{~b} K_{fcd} l^f ~.
\label{relnnablaalbnchi}
\end{equation}

The final kinematical quantity that will be of interest to us in our analysis is the transversal deformation rate tensor $\xih_{ab}$. This measures the projection (onto $S_t$) of the rate of change of the induced metric $q_{ab}$ as it evolves along the transverse auxiliary null vector field $\bm{k}$,
\begin{equation}
\hat{\Xi}_{ab} \equiv \fr q^c_{~a} q^d_{~b}~\lik q_{cd} ~.
\label{trandeform}
\end{equation}
Again, simply via the definition, it can be observed that the transversal deformation rate tensor is symmetric and orthogonal to $\bm{l}$ and $\bm{k}$. Performing an irreducible decomposition on the spatial bilinear $\xih_{ab}$, we see, 
\begin{equation}
\xih_{ab}  = \fr q_{ab} ~\thdk + \shdk ~.
\end{equation}
The trace $\thdk$ of the bilinear $\xih_{ab}$ is called the ingoing expansion scalar as it correctly measures the fractional rate of change of $\sqrt{q}$ as it evolves along $\bm{k}$. Similarly, $\shdk$ is the traceless shear part corresponding to $\xih_{ab}$. We would also find it in our interest to compute the spacetime covariant derivative of the auxiliary null field $\bm{k}$. To that end, it can seen \cite{Dey:2022qqp} that, 
\begin{equation}
\hcvd_a k_b = \xih_{ab} - \hhw_a k_b - k_a \hw_b -l_a (k^i \hcvd_i k_b) + k_a T_{cdf}k^c l^d q^f_{~a} - q^c_{~a} q^d_{~b} K_{fcd} k^f ~.
\label{nablaakbexpansion}
\end{equation}
\section{The gravitational ECKS field equations}\label{summary2}
In the Einstein-Cartan (EC) field theory, both the metric and the torsion tensor are treated as independent dynamical variables. The gravitational action (denoted by $\mathcal{A}_{\text{EC}}$) for this theory is simply \cite{Poplawski:2009fb, kleinert1989gauge},
\begin{align}
\mathcal{A}_{\text{EC}} =  \frac{1}{16 \pi} \int_{\mathcal{V}} d^4 x \sqrt{-g} \hr 
\end{align}
The dynamics of the metric is sourced by the energy-momentum tensor $\mt_{ab}$ whereas the dynamics of torsion is sourced by the spin angular momentum tensor $\tau^a_{~bc}$. The variation of the matter action is \cite{Poplawski:2009fb, kleinert1989gauge}, 
\begin{align}
\d \aem \equiv - \fr \int_{\mathcal{V}} d^4 x \sqrt{-g} \Big[\mt_{ab} \d g^{ab} + \tau^{b~c}_{~a}~ \d K^a_{~bc}\Big] ~.
\end{align} 
The resultant ECKS field equation obtained by varying the total action (gravitation plus matter) w.r.t the metric is \cite{Poplawski:2009fb, kleinert1989gauge},
\begin{align}
\hg_{ab}  +\fr (\hcvd_c + T_c) \Big(-S^c_{~ab} + S_{ab}^{~~c} + S_{ba}^{~~c}\Big) = 8 \pi \mt_{ab} ~.
\label{ECKSeom}
\end{align}
Extremizing the total action w.r.t the contorsion tensor leads to,
\begin{align}
S^a_{~bc} = 8 \pi \tau^a_{~bc} ~.
\label{spindensity}
\end{align}
Having equipped ourselves with all the necessary machineries, we are now in a position to proceed with our investigation.
\section{The dynamics pertaining to $\hg_{ab}l^a q^b_{~c}$}\label{dynamics1}
Now, that we have pointed out the relevant kinematical quantities of interest for our analysis, let us proceed to the dynamical variables. By ``kinematics'', we mean those quantities obtainable as the first order spacetime derivative of the null vector fields $\bm{l}$ and $\bm{k}$, the null $1$-forms $\underline{\bm{l}}$ and $\underline{\bm{k}}$ as well as the metric field $\bm{g}$ and $\bm{q}$ \cite{Gourgoulhon:2005ng}. By ``dynamics'', we mean the quantities involving first derivatives of kinematical objects along a specified vector field in the spacetime. The null generator $\bm{l}$ is related to the time evolution vector $\bm{t}$ and this is precisely why we would call the first derivative (Lie or covariant) of kinematical variables along $\bm{l}$ to be dynamical quantities of interest. Now, as far Einstein's theory of gravity is concerned, the field equations expressed w.r.t a generic null hypersurface $\hech$ give rise to important physical interpretations. To look at this, consider the expansion of the vector field $G^a_{~b}l^b$ in the basis $(\bm{l}, \bm{k}, \bm{\p_A})$,
\begin{equation}
G^a_{~b} l^b = \phi_1 l^a + \phi_2 k^a + \phi^{{A}} (\bm{\p_A})^{a} ~,
\end{equation}
with $\phi_1 = -G_{ab} k^al^b$, $\phi_2 = -G_{ab}l^a l^b$ and $\phi_{{A}} = (G_{ab}l^b q^a_{~c}) e^c_{~{A}}$. As has been mentioned in the introduction, these projection components have important thermodynamic and fluid-dynamic interpretation as far as Einstein gravity is concerned. 
It would be worthwhile to investigate whether these physical notions carry forward to other theories of gravity as well. It is in this context, that we are investigating the structure of the ECKS field equation w.r.t to a generic null hypersurface $\hech$ in the RC spacetime. The corresponding NRE for the EC theory (determined through $-\phi_2$) and its physical interpretation has already been presented in \cite{Dey:2017fld, Dey:2022qqp} and will not be pursued here. The thermodynamic interpretation provided to the ECKS field equations via the component $\hg_{ab}k^a l^b$ has already been presented by the authors in \cite{Dey:2022qqp}. In this work, we will try to establish and clarify the dynamics of the Hajicek $1$-form $\hhw_a$. We will indeed see that dynamical evolution of $\hhw_a$ along $\bm{l}$ is related to the projection component $\hg_{ab}l^a q^b_{~c}$. Then, we will try to answer the question as to whether it is even possible to interpret the resulting equation as some kind of (modified) DNS equation. 

So let us look at the dynamics of the generic null hypersurface via the projection component $\hg_{ab}l^aq^b_{~c}$. Let us begin with the null Ricci identity established for $l^a$,
\begin{align}
\Big[\hcvd_b ,\hcvd_a\Big]l^b = \hr_{ca}l^c - T^i_{~ba}(\hcvd_i l^b) ~.\label{Ricciiden}
\end{align}
We will be interested in the projection of the above dynamical equation on the transverse two surface $S_t$ of $\hech$,
\begin{align}
\Big[\hcvd_b ,\hcvd_a\Big]l^b q^a_{~t} = \hr_{ab}l^a q^b_{~t} - T^i_{~ba}(\hcvd_i l^b)q^a_{~t} ~.
\label{projRicciiden}
\end{align} 
From here along, our procedure will follow \cite{Gourgoulhon:2005ng}. In order to remove the clutter of indices, let us define the following spatial tensor,
\begin{align}
\tit_{ij} \equiv q_j^{~r} q_i^{~s}K_{trs}l^t ~.
\label{tit}
\end{align}
 Let us also define the following quantity,
\begin{align}
\hat{\Phi}^b_{~a} \equiv \hchi^b_{~a} - q^c_{~a} q^{db} K_{fcd} l^f = \hchi^b_{~a} - \tit^b_{~a} ~.
\label{hphi}
\end{align}
$\hphi^b_{~a}$ is a completely spatial $(1,1)$ tensor that is orthogonal to both the $\bm{l}$ and $\bm{k}$ directions. It is easily verified that,
\begin{align}
\hphi_{ab} - \hphi_{ba} = 2 \hphi_{[ab]} = q^c_{~a} q^d_{~b} T_{fcd}l^f ~.\label{manip4}
\end{align}

It can be shown that the (Lie) evolution of the Hajicek $1$-form $\underline{\bm{\hhw}}$ along the null generator $\bm{l}$ is indeed related to the quantity $\hr_{ab}l^a q^b_{~t} = \hg_{ab}l^a q^b_{~t}$ via the following relationship,
\begin{align}
& q^a_{~t} \lil \hhw_a + \hsvd_b \hphi^b_{~t} - \hchi_{bt} \hp^b + q^c_{~t} q^d_{~b}(K_{fcd}l^f) \hp^b - \xih_{tb} \ti^b \nonumber \\
& + \hhw_t \Big(\thdl + \ktb - T_bl^b\Big) - \hsvd_t \Big(\thdl + \kappa -T_b l^b\Big) + q^c_{~t} q^d_{~b}(K_{fdc}k^f) \ti^b \nonumber \\
& = \hr_{ab}l^a q^b_{~t} - T_{iba} \hchi^{bi} q^a_{~t} + q^a_{~t}(T_{iba}k^i q^{cb} \ti_c) + q^a_{~t}(q^{ci}q^{db}T_{iba}(K_{fcd}l^f)) ~,
\label{evo1}
\end{align}
where $\bm{\hsvd}$ is the spatial covariant derivative compatible with the induced metric of the submanifold $(S_t, \bm{q})$,  i.e $\hsvd_a q_{bc} = 0$. 
The explicit proof of the above equation \eqref{evo1} has been provided in Appendix \ref{evol1}.
We now consider the completely spatial tensor $\hphi_{bt}$ and perform an irreducible decomposition of it by breaking it up into a trace part ($\overset{*}{\theta}$), a symmetric traceless part ($\overset{*}{\sigma}_{bt}$) and an anti-symmetric traceless part (${\overset{*}{\omega}}_{bt}$). The trace of this spatial tensor $\hphi_{bt}$ is,
\begin{align}
\stth &= q^{bt}\hphi_{bt} = q^bt \Big(\hchi_{bt} - q^c_{~t}q^d_{~b}K_{fcd}l^f\Big) = \thdl - q^{dc} K_{fcd}l^f \nonumber \\
& = \thdl - (g^{cd} + l^c k^d + k^cl^d)K_{fcd}l^f = \thdl - T_b l^b + \ti_b k^b ~.
\label{sttheta}
\end{align}
In arriving at the above result, we have used the antisymmetry of the contorsion tensor in the first and third indices along with the fact that $K_{a~~b}^{~b} = T_a$. The symmetric traceless part of the tensor $\hphi_{bt}$ is ,
\begin{align}
\stsig_{bt} &= \hphi_{(bt)} - \fr q_{bt} \stth = \Big(\hchi_{(bt)} - q^c_{~t}q^d_{~b}K_{f(cd)}l^f\Big) - \fr q_{bt} \Big(\thdl - T_b l^b + \ti_b k^b\Big) \nonumber \\
& = \Big(\hchi_{bt} - \fr q_{bt} \thdl\Big) - \Big(q^c_{~t}q^d_{~b} K_{f(cd)}l^f - \fr q_{bt}(T_b l^b - \ti_b k^b)\Big) \nonumber \\
& = {^{(\bm{l},d)}{\sigma}}_{bt} - \Big(q^c_{~t}q^d_{~b} K_{f(cd)}l^f - \fr q_{bt}(T_b l^b - \ti_b k^b)\Big) ~.
\label{stsigma}
\end{align}
Similarly, the traceless anti-symmetric part of $\hphi_{bt}$ is,
\begin{align}
\stome_{bt} = \hphi_{[bt]} = \hchi_{[bt]} - \fr q^c_{~t} q^d_{~b} (K_{fcd} - K_{fdc})l^f = \fr q^c_{~t}q^d_{~b}T_{fdc}l^f ~.
\label{stomega}
\end{align}
Using \eqref{sttheta}, \eqref{stsigma} and \eqref{stomega}, we obtain,
\begin{align}
\hphi_{bt} = \fr q_{bt} \Big(\thdl - T_b l^b + \ti_b k^b\Big) + \stsig_{bt} + \fr q^c_{~t} q^d_{~b} T_{fdc}l^f ~.
\label{irrephphi}
\end{align}
The above result allows us to have,
\begin{align}
\hsvd_b \hphi^b_{~t} = \hsvd_t \Big[\fr\Big(\thdl - T_b l^b + \ti_b k^b\Big)\Big] + \hsvd_b \stsig^b_{~t} + \hsvd^b \Big(\fr q^c_{~t} q^d_{~b}T_{fdc}l^f\Big) ~.\label{manip13}
\end{align}
Using \eqref{manip13} in \eqref{evo1} and further simplifying, we end up with,
\begin{align}
& q^a_{~t} \lil \hhw_a + \hhw_t \Big(\thdl - T_b l^b + \ktb\Big) - \hsvd_t \Big(\kappa + \fr (\thdl - T_b l^b - \ktb)\Big) + \hsvd_b \stsig^b_{~t} \nonumber \\
& + \hsvd^b \Big(\fr q^c_{~t} q^d_{~b}T_{fdc}l^f\Big) - \hchi_{bt} \hp^b + q^c_{~t}q^d_{~b} (K_{fcd}l^f) \hp^b - \xih_{tb} \ti^b + q^c_{~t}q^d_{~b}(K_{fdc}k^f) \ti^b \nonumber \\
& = \hr_{ab}l^a q^b_{~t} - T_{iba} \hchi^{bi}q^a_{~t} + q^a_{~t}(T_{iba}k^i q^{cb} \ti_c) + q^a_{~t}(q^{ci}q^{db}T_{iba}(K_{fcd}l^f)) ~.
\label{evo2}
\end{align}
The above geometrical relationship established in the RC spacetime relates the (Lie)evolution of the Hajicek one-form $\hhw_a$ along the null generators in its full generality with the projection component $\hr_{ab}l^a q^b_{~t}$ or $\hg_{ab}l^a q^b_{~t}$. Note that, in arriving at \eqref{evo2}, we have not used the geodesic constraint condition $\ti_b = 0$. 

We can express the above relationship in a slightly different version by switching over to the extended second fundamental form $\thht_{bt}$. This we do by noticing via the relationship between the extended second fundamental form and the deformation rate tensor, i.e \eqref{relnchintheta}. We see that upon using \eqref{relnchintheta}, we have,
\begin{align}
-\thht_{ba}\hp^b q^a_{~t} = - \hchi_{bt} \hp^b + q^c_{~t}q^d_{~b} (K_{fcd}l^f) \hp^b ~.
\label{manip14}
\end{align} 
Similarly the last three terms on the R.H.S of \eqref{evo2} can be combined via \eqref{relnchintheta} to get,
\begin{align}
- T_{iba} \hchi^{bi}q^a_{~t} + q^a_{~t}(T_{iba}k^i q^{cb} \ti_c) + q^a_{~t}(q^{ci}q^{db}T_{iba}(K_{fcd}l^f)) &= -T_{iba}q^a_{~t} \Big(\hchi^{bi} - k^i q^{cb} \ti_c - q^{ci}q^{db}K_{fcd}l^f\Big) \nonumber \\
& = -T_{iba}q^a_{~t} \thht^{bi} ~.
\label{manip15}
\end{align}
Using \eqref{manip14} and \eqref{manip15} in \eqref{evo2}, we express the Lie-evolution of the Hajicek one-form in a slightly different form,
\begin{align}
& q^a_{~t} \lil \hhw_a + \hhw_t \Big(\thdl - T_b l^b + \ktb\Big) - \hsvd_t \Big(\kappa + \fr (\thdl - T_b l^b - \ktb)\Big) + \hsvd_b \stsig^b_{~t} \nonumber \\
& + \hsvd^b \Big(\fr q^c_{~t} q^d_{~b}T_{fdc}l^f\Big) - \thht_{ba}\hp^b q^a_{~t} - \xih_{tb} \ti^b + q^c_{~t}q^d_{~b}(K_{fdc}k^f) \ti^b \nonumber \\
& = \hr_{ab}l^a q^b_{~t} - T_{iba}\thht^{bi}q^a_{~t} ~.
\label{evo3}
\end{align}
Note that, in contrast to the deformation rate tensor, the extended second fundamental form is not symmetric and completely spatial as evidenced from the relations \eqref{extfundformprops} and \eqref{diffinTheta}. 

Next, we have to use the gravitational field equations to interpret \eqref{evo2} or \eqref{evo3} as a dynamical evolution equation of the Hajicek one-form $\hhw_a$. To this extent, we will focus exclusively on the EC theory, under which the relevant equations are the ECKS field equations \eqref{ECKSeom}. We then have,
\begin{align}
\hr_{ab}l^a q^b_{~t} = \hg_{ab}l^a q^b_{~t} = 8 \pi \mt_{ab} l^a q^b_{~t} - \fr (\hcvd_c +T_c) \Big(-S^c_{~ab}+ S_{ab}^{~~c} + S_{ba}^{~~c}\Big)l^a q^b_{~t} ~.
\label{gablaqbt} 
\end{align} 
Note that in \eqref{evo3} or \eqref{evo2}, we have related the dynamical evolution (along $\bm{l}$) of $\underline{\bm{\hhw}}$ with $\hr_{ab}l^a q^b_{~t} = \hg_{ab}l^a q^b_{~t} $ instead of $\hr_{ab}q^a_{~t} l^b = \hg_{ab}q^a_{~t}l^b$ as anticipated in the transverse projection component of the vector field $\hg^a_{~b}l^b$. However, this is not a matter of concern, since at this point, we eventually use the field equations as seen in \eqref{gablaqbt}.
Using \eqref{gablaqbt} in \eqref{evo3}, we have,
\begin{align}
& q^a_{~t} \lil \hhw_a + \hhw_t \Big(\thdl - T_b l^b + \ktb\Big) - \hsvd_t \Big(\kappa + \fr (\thdl - T_b l^b - \ktb)\Big) + \hsvd_b \stsig^b_{~t} \nonumber \\
& + \hsvd^b \Big(\fr q^c_{~t} q^d_{~b}T_{fdc}l^f\Big) - \thht_{ba}\hp^b q^a_{~t} - \xih_{tb} \ti^b + q^c_{~t}q^d_{~b}(K_{fdc}k^f) \ti^b \nonumber \\
& = 8 \pi \mt_{ab} l^a q^b_{~t} - \fr (\hcvd_c +T_c) \Big(-S^c_{~ab}+ S_{ab}^{~~c} + S_{ba}^{~~c}\Big)l^a q^b_{~t} - T_{iba}\thht^{bi}q^a_{~t} ~.
\label{devo1}
\end{align}
The above relation reduces to the equation $(6.15)$ in \cite{Gourgoulhon:2005ng} in the absence of torsion in the spacetime and then hence defines the Hajicek equation under the membrane paradigm.
We notice that the above general dynamical evolution law of the Hajicek one-form $\hhw_a$ i.e \eqref{devo1} involves the transversal deformation rate tensor $\xih_{tb}$. The transversal deformation rate tensor in essence measures the projection of the Lie derivative of the transverse induced metric $q_{ab}$ along the ingoing auxiliary null field $\bm{k}$. So the dynamics involving the evolution of the Hajicek one-form i.e \eqref{devo1} involves a part that contains the evolution of $q_{ab}$ along $\bm{k}$. This is however in stark contrast to the evolution equation of the Hajicek one-form in the Einstein spacetime \cite{Gourgoulhon:2005ng}. Such a dynamical evolution equation in the Einstein spacetime involves the evolution of relevant kinematical quantities along the outgoing null generator $\bm{l}$ and none so in the direction of the ingoing field $\bm{k}$. This can also be explicitly seen from the corresponding NRE in the RC spacetime. Without invoking the geodesic constraint, the most general form of the NRE has been derived in \cite{Dey:2022qqp} (see Eq. $(A.18)$ of \cite{Dey:2022qqp}). A different variant of the NRE in RC spacetime has been derived in the Appendix \ref{tidal} (see Eq. \eqref{thdlevol}). It is quite clear that in the presence of torsion in the spacetime, the dynamical evolution of the outgoing expansion scalar $\thdl$ along $\bm{l}$ encodes information about the uniquely defined ingoing $\bm{k}$ field. This is again in contrast with the Einstein spacetime, where the NRE carries no information about the auxiliary $\bm{k}$ field. Coming back to the present analysis, in the presence of torsion in the spacetime, i.e in the RC spacetime, the evolution equation of the Hajicek $1$-form $\hhw_a$ involves terms like $\xih_{tb}$. This is preferably due to the fact that in the RC spacetime, the null generators of our integrable null hypersurface $\hech$ are not parallel transported along themselves. However these null generators are themselves null geodesics w.r.t to the Levi-Civita connection $\bm{\nabla}$. This is quite evident from the relation \eqref{parrl1}. However, if we impose the geodesic constraint i.e $\ti_b = T_{abc}l^a l^c = 0$, then we force the null generators to be simultaneously auto-parallel (w.r.t to the connection $\bm{\hat{\nabla}}$) and extremal length geodesics (w.r.t $\bm{\nabla}$). Hence, we notice that under the geodesic constraint, we remove in equations \eqref{evo2} and \eqref{evo3} any reference to evolution of kinematical/geometrical quantities along the auxiliary null field $\bm{k}$. As a side note, this feature was also shared by the NRE in the RC spacetime, where the application of the geodesic constraint removed explicit references of evolution of kinematical quantities along $\bm{k}$. Thus, under the geodesic constraint, we have our relevant dynamical evolution laws for $\hhw_a$ to be,
\begin{align}
& q^a_{~t} \lil \hhw_a + \hhw_t \Big(\thdl - T_b l^b\Big) - \hsvd_t \Big(\kappa + \fr (\thdl - T_b l^b )\Big) + \hsvd_b \stsig^b_{~t} + \hsvd^b \Big(\fr q^c_{~t} q^d_{~b}T_{fdc}l^f\Big)  \nonumber \\
&- \thht_{ba}\hp^b q^a_{~t} = 8 \pi \mt_{ab} l^a q^b_{~t} - \fr (\hcvd_c +T_c) \Big(-S^c_{~ab}+ S_{ab}^{~~c} + S_{ba}^{~~c}\Big)l^a q^b_{~t} - T_{iba}\thht^{bi}q^a_{~t} ~.
\label{lieomega1} 
\end{align}
The above can be expressed in the following alternative structure as well:
\begin{align}
& q^a_{~t} \lil \hhw_a + \hhw_t \Big(\thdl - T_b l^b \Big) - \hsvd_t \Big(\kappa + \fr (\thdl - T_b l^b )\Big) + \hsvd_b \stsig^b_{~t} \nonumber \\
& + \hsvd^b \Big(\fr q^c_{~t} q^d_{~b}T_{fdc}l^f\Big) - \hchi_{bt} \hp^b + q^c_{~t}q^d_{~b} (K_{fcd}l^f) \hp^b  \nonumber \\
& = 8 \pi \mt_{ab} l^a q^b_{~t} - \fr (\hcvd_c +T_c) \Big(-S^c_{~ab}+ S_{ab}^{~~c} + S_{ba}^{~~c}\Big)l^a q^b_{~t} \nonumber \\
&- T_{iba} \hchi^{bi}q^a_{~t}  + q^a_{~t}(q^{ci}q^{db}T_{iba}(K_{fcd}l^f)) ~.
\label{lieomega2}
\end{align}  

The above dynamical equations \eqref{lieomega1} and \eqref{lieomega2} are completely covariant relations written in any arbitrary coordinate system. It is this equation that in absence of torsion in the spacetime reduces to the Damour-Navier-Stokes (DNS) equation. So, in view of this, let us discuss the structural aspects of this dynamical equation. Firstly, note that \eqref{lieomega1} contains the Lie derivative of the Hajicek $1$-form along the null generator $\bm{l}$. In a coordinate system adapted to the null hypersurface $\hech$, we have \cite{Gourgoulhon:2005ng}, 
\begin{align}
\bm{l} \overset{\hech}{=} \bm{t} + \bm{V} ~,
\label{adapted}
\end{align}
where $\bm{t}$ is the time evolution vector field (essentially connecting similar spatial points on the   $t=\text{constant}$ spacelike $\Sigma_t$ slices foliating $\hech$) and $\bm{V}$ is a spacelike vector field tangent to the two-surface $S_t$. For a coordinate system $(t,x^{\mu} = \{x^1, x^2, x^3\})$ adapted to the null hypersurface $\hech$, its location is prescribed by say $x^1 = 1$ and hence the coordinates on the transverse $2$-surface are prescribed by $(x^A = \{x^2,x^3\})$. For such an adapted coordinate system, we hence have $l^a \overset{\hech}{=} t^a + V^A \bm{\partial}_A$ and that $q^a_{~A} = \delta^a_{~A}$. For such a system, its quite easy to show that,
\begin{align}
q^a_{~A} \lil \hhw_a \overset{\hech}{=}q^a_{~A} \pounds_{\bm{t} + \bm{V}} \hhw_a = \frac{\p \hhw_{A}}{\p t} + V^B \hsvd_B \hhw_A + \hhw_B \hsvd_A V^B + {^{(2)}}T^B_{~CA}V^C \hhw_B ~,
\label{adapted1}
\end{align}
where ${^{(2)}}T^A_{~BC}$ is the induced torsion tensor on the transverse submanifold $(S_t, \bm{q}, \bm{\hsvd})$. Inserting the above relation \eqref{adapted1} into \eqref{lieomega1}, we obtain,
\begin{align}
&\frac{\p \hhw_{A}}{\p t} + V^B \hsvd_B \hhw_A + \hhw_B \hsvd_A V^B + {^{(2)}}T^B_{~CA}V^C \hhw_B + \hhw_A (\thdl - T_b l^b) \nonumber \\
& = 8 \pi q^b_{~A}\mt_{ab}l^a - \fr (\hcvd_c +T_c) \Big(-S^c_{~ab}+ S_{ab}^{~~c} + S_{ba}^{~~c}\Big)l^a q^b_{~A}+ \thht_{ba}\hp^b q^a_{~A} - T_{iba} \thht^{bi} q^a_{A}  \nonumber \\
&+ \hsvd_A \kappa- \hsvd_B \stsig^B_{~A} + \fr \hsvd_A (\thdl - T_b l^b) - \hsvd^B \Big(\fr q^c_{~A} q^d_{~B}T_{fdc}l^f\Big)~.
\label{convective1}
\end{align} 
The initial two terms in the L.H.S of \eqref{convective1} denote the material derivative of the Hajicek $1$-form $\hhw_A$ along $\bm{V}$. In the context of Einstein gravity, as applied to a black-hole event horizon, Damour interpreted $\bm{V}$ as the surface velocity of the horizon. In this context, $\bm{V}$ is to be interpreted as the surface velocity of the null hypersurface $\hech$ w.r.t the adapted coordinates in the RC spacetime. The extra term $\hhw_B \hsvd_A V^B$ involving the derivative of the velocity field $V^A$ is present (as in the DNS case) along with an extra term ${^{(2)}}T^B_{~CA}V^C \hhw_B$. This is because the ambient spacetime $\sptm$ carries intrinsic torsion that induces a non-symmetric connection $\bm{\hsvd}$ and hence torsion on the submanifold $(S_t, \bm{q}, \bm{\hsvd})$. 
It is worth mentioning that in analogy with the \textit{membrane paradigm} approach, we can exchange the Lie derivative operator with the operator $\hat{\mathcal{D}}_{\bar{t}}$ \cite{PhysRevD.33.915} such that its operation on the Hajicek $1$-form is given as, 
\begin{align}
\spdt \hhw_i \equiv q^a_{~i}(l^j \hcvd_j \hhw_a) ~,
\label{spdt}
\end{align}
which quantifies the projection (onto $S_t$) of the covariant derivative of the Hajicek $1$-form along $\bm{l}$. It can then quite easily be seen that,
\begin{align}
q^a_{~i} \lil \hhw_a = \spdt \hhw_i + \hhw_k \thht^k_{~j}q^j_{~i} + T_{kja}l^j \hhw^k q^a_{~i} ~.
\label{spdtlie}
\end{align}
Using \eqref{spdtlie} in \eqref{lieomega1}, we can also interpret the dynamical evolution of the Hajicek $1$-form in the following way,
\begin{align}
&\spdt \hhw_i  + \hhw_i \Big(\thdl - T_b l^b\Big) - \hsvd_i \Big(\kappa + \fr (\thdl - T_b l^b )\Big) + \hsvd_j \stsig^j_{~i} + \hsvd^j \Big(\fr q^c_{~i} q^d_{~j}T_{fdc}l^f\Big) + (\hhw_k -\hp_k) \thht^k_{~i}  \nonumber \\
& = 8 \pi \mt_{jk} l^j q^k_{~i} - \fr (\hcvd_j +T_j) \Big(-S^j_{~mn}+ S_{mn}^{~~~j} + S_{nm}^{~~~j}\Big)l^m q^n_{~i} - T_{kja}(\thht^{jk}+ l^j \hhw^k)q^a_{~i} ~.
\label{hajicekeqn}
\end{align} 
In the absence of torsion in the spacetime, the above equation reduces to the ``Hajicek equation'' arising in the context of membrane paradigm approach \cite{PhysRevD.33.915}. Though we have these two notions of the derivative operator acting on $\hhw_a$, the Lie derivative presents itself as a natural generalization to the material derivative in curved manifolds. 
\section{Possible connections with fluid dynamics/ generalized continuum mechanics}\label{interpretation}
\subsection{Projected ECKS field equations on $\hech$ are not equivalent to DNS :}\label{interpretation1}
It is quite well known that for the case of vanishing torsion, the DNS equation describes the dynamics of a $2$-dimensional null viscous fluid living on $\hech$. In fact, it has also been shown that the DNS equation is exactly identical to the NS equation provided we view it in a boosted inertial frame \cite{Padmanabhan:2010rp}. Much in the same way, we would like to have an interpretation for the dynamical equations \eqref{lieomega1} or \eqref{convective1}.

We notice in \eqref{lieomega1}, the presence of the spatial covariant derivatives of $\stsig_{ab}$ and $(\fr q^c_{~a}q^d_{~b}T_{fdc}l^f)$ which denote the traceless symmetric and antisymmetric parts of $\hphi_{ab}$ respectively. With respect to the adapted coordinate system $(x^a = (t, x^{\mu}) = (x^0, x^{\mu}))$, it can be shown that, (for derivation, see Appendix \ref{adapx})
\begin{align}
\hchi_{AB} \overset{\hech}{=} \fr\Big(\p_t q_{AB} + \hsvd_A V_B + \hsvd_B V_A + ({^{(2)}}T_{ACB}+ {^{(2)}}T_{BCA})V^C\Big) ~,
\label{adapchi}
\end{align}
and
\begin{align}
\hphi_{AB} = \hchi_{AB} - \tit_{AB} \overset{\hech}{=} \fr\Big(\p_t q_{AB} + \hsvd_A V_B + \hsvd_B V_A - 2 K_{0BA} - {^{(2)}}T_{DBA}V^D\Big)~.
\label{adapphi}
\end{align}
In the case of the geodesic constraint ($\ti_a = 0$), it can easily be verified that $\thht_{ab} \overset{\ti_a =0}{ = } \hchi_{ab} - \tit_{ab} = \hphi_{ab}$ i.e the second fundamental form of the null hypersurface $\hech$ coincides with the spatial tensor $\hphi_{ab}$. Hence we have,
\begin{align}
\thht_{BA} \overset{\ti_a =0}{ = } \hphi_{BA} = \hchi_{BA} - \tit_{BA} \overset{\hech}{=} \fr\Big(\p_t q_{AB} + \hsvd_A V_B + \hsvd_B V_A - 2K_{0AB} - {^{(2)}}T_{DAB}V^D\Big)~.
\label{adapth}
\end{align} 
We do indeed see that w.r.t the adapted coordinate system, $\thht_{BA}$ contains the term $\hsvd_A V_B + \hsvd_B V_A$. Let us mention the reason as to why we are looking at the spatial tensor $\thht_{BA}$. In the absence of torsion, for the case of Einstein gravity, the second fundamental form of $\hech$ in $(\mathcal{M}, \bm{g}, \bm{\nabla})$ is of the form,
\begin{align}
\Theta_{BA} \overset{(\mathcal{M}, \bm{g}, \bm{\nabla})}{=} \fr( \p_t q_{AB} + \mathcal{D}_A V_B + \mathcal{D}_B V_A) ~.
\label{manip34}
\end{align}
For the suitable choice case of an adapted coordinate system on the null surface $\hech$, we can make the induced metric $q_{AB}$ of $S_t$ independent of the time evolution parameter $t$ i.e $\p_t q_{AB} = 0$. Then, we notice that the second fundamental form of $\hech$ in Einstein gravity has the same form as that of the stress tensor of a viscous fluid (having no internal angular momentum) with velocity $V_A$. The symmetric combination of the velocity gradient tensor i.e $1/2 (\mathcal{D}_A V_B + \mathcal{D}_B V_A)$ can as usual be broken down into a trace part and a traceless shear part. The trace part which contains the divergence of the velocity field $V_A$ is necessarily interpreted as the expansion scalar corresponding to the fluid flow lines. In that same respect, the trace of the second fundamental form $\T_{AB} = \T_{BA}$ gives the true expansion scalar of the null surface $\hech$. This is perhaps the central reason as to why (in the absence of torsion), \eqref{lieomega1} or \eqref{convective1} can be interpreted as the DNS equation or the NS equation (in a boosted inertial frame) \cite{Padmanabhan:2010rp}.
The viscous stress tensor for a conventional  two dimensional NS fluid is necessarily of the form $2 \eta \sigma_{AB} + \xi \delta_{AB} \theta$, where $\eta$ and $\xi$ stand for the shear and bulk viscosity coefficients respectively. For the NS fluid, the trace free shear tensor $\sigma_{AB}$ is built from the derivatives of the velocity field $V_A$. In the case of vanishing torsion tensor, the spatial tensors $\thht_{AB}$, $\hphi_{AB}$ and $\hchi_{AB}$ all coincide.

 Coming back to the EC theory, we indeed see via \eqref{adapth} that the second fundamental form of $\hech$ as usual contains the term $\hsvd_A V_B + \hsvd_B V_A$. 
 However, now for the generic RC spacetime $\sptm$, $\thht_{BA}$ has the extra terms of $\p_t q_{AB}$  and $- 2 K_{0AB} - {^{(2)}}T_{DAB}V^D$ involving the contorsion and the two dimension torsion tensor. These terms have no direct interpretation of fluid variables. Via certain choices of the metric and the adapted coordinate system along with the freedom of rescaling $\bm{l}$, one can make $q_{AB}$ independent of t. However, the term involving the contorsion and torsion tensor i.e $- 2 K_{0AB} - {^{(2)}}T_{DAB}V^D$ cannot be set to zero for the generic RC spacetime. Moreover, now in the RC spacetime, the second fundamental form $\thht_{BA}$ has both symmetric as well as antisymmetric parts due to the presence of the torsion tensor in its definition \eqref{adapth}. This is certainly a far cry from the usual spacetime without torsion. It is precisely for this reason that by a particular choice of the metric and an adapted coordinate system, we cannot set $\p_t q_{AB} - 2 K_{0AB} - {^{(2)}}T_{DAB}V^D = 0$. It will not be possible to compensate for the antisymmetric contribution. 
 The presence of the antisymmetric part in the definition \eqref{adapth} of $\thht_{BA}$ is the reason as to why we have a term $ \hsvd^b \Big(\fr q^c_{~t} q^d_{~b}T_{fdc}l^f\Big)$ involving the spatial derivative of an antisymmetric tensor. This shows that in no way, can \eqref{lieomega1} or \eqref{convective1} be interpreted as some kind of (modified) DNS equation in the RC spacetime. This is because the stress tensor for the DNS fluid is by default symmetric. The precise symmetry property of the Cauchy stress tensor is intimately connected to the underlying assumption that the material points describing the continuum (fluid) system have no intrinsic angular momentum. But this does fly in the face of the assumption underlying the RC spacetime in the EC theory. The very source of torsion is the intrinsic spin-angular momentum tensor describing the intrinsic spin of particles in the geometrization of RC spacetime. Thus there is a non-zero contribution involving the spatial covariant derivative of a completely antisymmetric tensor in \eqref{lieomega1} and \eqref{convective1}. It is for this reason, we can safely conclude that if we were to allude a fluid/elastic medium interpretation to the dynamics of our null hypersurface $\hech$ in EC theory, its shear tensor would certainly not be symmetric in general. Hence it is quite certain, that the projected ECKS field equations on $\hech$ (as observed via the projection component $\hg_{ab}l^a q^b_{~c}$) cannot be interpreted as the DNS (or NS) fluid equation.

\subsection{Any real life analogies possible?}\label{interpretation2}
Perhaps the most that we can stretch our imagination to give a real life analogy with the dynamics of \eqref{lieomega1} or \eqref{convective1} is that of Cosserat \cite{vardoulakis2019cosserat} fluids. In fact, Cosserat theory describes a classical elastic continuum in which the material point bodies have translation as well as rotational degrees of freedom. Conventional Navier-Stokes dynamics does not incorporate any intrinsic length scales. Each point of the Cosserat medium can be visualized as an infinitesimal rigid body. There exists both stress and couple stress as responses to the translation and rotational degrees of freedom respectively. 

Let us now briefly describe the Cosserat fluid in Cartesian coordinates $x^{\mu}$. The particle velocity and spin are designated as $v_{\mu}(t,x^{\nu})$ and $w_{\mu}(t,x^{\nu})$ respectively. The velocity gradient tensor can as usual be broken into a symmetric and antisymmetric part,
\begin{align}
\p_{\mu} v_{\nu} = D_{\mu \nu} + W_{\mu \nu} ~,
\label{cos1}
\end{align} 
where $D_{\mu \nu} = \fr (\p_{\mu} v_{\nu} + \p_{\nu}v_{\mu})$ and $W_{\mu \nu} = \fr (\p_{\mu} v_{\nu} - \p_{\nu}v_{\mu})$ represent strain rate (or classical deformation rate) tensor and the rotation rate (or vorticity) tensor respectively. The Cosserat generalization to the NS equations are then described as \cite{vardoulakis2019cosserat},
\begin{align}
\rho \Big(\p_t v_{\alpha} + v^{\mu} \p_{\mu} v_{\alpha}\Big) = \p_{\a} P + \g \p^{\mu}(\p_{\mu} v_{\a} + \p_{\a} v_{\mu}) + \g_c \p^{\mu}([\p_{\mu} v_{\a} - \p_{\a} v_{\mu}] - 2 \epsilon_{\mu \alpha \rho} w ^{\rho}) + f^{\text{ext}}_{\alpha} ~.
\label{cos2}
\end{align}
In the above Eq. \eqref{cos2}, $P$ stands for the fluid pressure and $\g$ is identified as the classical macroscopic fluid viscosity and is related to the symmetric part of the shear tensor. The quantity $\g_c$ represents an extra viscosity material parameter. The relative spin of the Cosserat fluid particle w.r.t the background vorticity is accounted for by the extra viscosity parameter $\g_c$ \cite{vardoulakis2019cosserat}. $f^{\text{ext}}_{\alpha}$ represents the external volume or body force density that acts per unit volume of the fluid.

\subsubsection{In coordinates adapted to $\hech$}
Let us rewrite \eqref{convective1} in the following suggestive way,
\begin{align}
&\p_t\Big(\frac{-\hhw_{A}}{8 \pi}\Big) + V^B \hsvd_B \Big(\frac{-\hhw_{A}}{8 \pi}\Big) +\Big(\frac{-\hhw_{B}}{8 \pi}\Big) \hsvd_A V^B + {^{(2)}}T^B_{~CA}V^C \Big(\frac{-\hhw_{B}}{8 \pi}\Big) + \Big(\frac{-\hhw_{A}}{8 \pi}\Big) (\thdl - T_b l^b) \nonumber \\
& = - q^b_{~A}\mt_{ab}l^a + \frac{1}{16 \pi} (\hcvd_c +T_c) \Big(-S^c_{~ab}+ S_{ab}^{~~c} + S_{ba}^{~~c}\Big)l^a q^b_{~A} - \frac{1}{8 \pi} \thht_{ba}\hp^b q^a_{~A} + \frac{1}{8 \pi} T_{iba} \thht^{bi} q^a_{A}  \nonumber \\
&- \hsvd_A \Big(\frac{\k}{8 \pi}\Big) + \hsvd^B \Big[\frac{1}{ 8 \pi} \Big(\stsig_{BA} - \fr q_{AB}(\thdl - T_b l^b)\Big)\Big] + \hsvd^B \Big[\frac{1}{8 \pi}\Big(-\fr q^d_{~B} q^c_{~A}T_{fcd}l^f\Big)\Big]~.
\label{convective4}
\end{align}
Now referring to \eqref{convective4}, we make the following identification,
\begin{align}
f^{\text{ext}}_{A} \equiv -\mt_{ab} q^b_{~A}l^a + \frac{1}{16 \pi} (\hcvd_c +T_c) \Big(-S^c_{~ab}+ S_{ab}^{~~c} + S_{ba}^{~~c}\Big)l^a q^b_{~A} - \frac{1}{8 \pi}\Big(\thht_{ba}\hp^b q^a_{~A} -T_{iba} \thht^{bi} q^a_{A}\Big) ~,
\label{cos3}
\end{align}
where $f^{\text{ext}}_{A}$ represents the surface force density on the two dimensional cross section $S_t$. It is the momentum per unit area per unit coordinate time $t$. We notice that the definition of force density includes the projection of matter energy momentum tensor onto the two surface $S_t$ via $-\mt_{ab} q^b_{~A}l^a$. However, it also includes contribution due to the torsion tensor as well. Specifically we note that in this identification that we have made for the force density, there is the existence of the terms $\thht_{ba}\hp^b q^a_{~A}$ and $-T_{iba} \thht^{bi} q^a_{A}$. This means that the torsion tensor coupled with the second fundamental form $\thht_{ab}$ provides a kind of a force density on the surface $S_t$. Once such identification has been made, we make comparisons between \eqref{convective4} and \eqref{cos2}. This will allow us to make the necessary identification for the description of the two dimensional viscous null fluid governed by \eqref{convective4}. The surface momentum density $\pi_A$ is related to the Hajicek $1$-form as 
\begin{align}
\pi_A = -\frac{1}{8 \pi} \hhw_A ~.\label{momden}
\end{align}
The velocity of our null fluid is $V^A$, whereas $\kappa/ (8 \pi)$ denotes the null fluid pressure. From the symmetric part of the stress tensor, we note that $1/(16 \pi)$ is the shear viscosity coefficient, whereas $-1/(16 \pi)$ represents the bulk viscosity coefficient. When comparing \eqref{cos2} and \eqref{convective4}, we notice that the comparative terms (for the antisymmetric part of the stress tensor) are respectively, $\gamma_c \p^{\mu}(\p_{\mu} v_{\alpha} - \p_{\alpha} v_{\mu} - 2 \epsilon_{\mu \alpha \rho} w^{\rho})$ and $\hsvd^B \Big[\frac{1}{8 \pi}\Big(-\fr q^d_{~B} q^c_{~A}T_{fcd}l^f\Big)\Big]$. It is clear that on the side of the null fluid dynamics \eqref{convective4}, there does not exist any term of the form $\hsvd^B(\hsvd_B V_A - \hsvd_A V_B)$. This is also evident from the expansion the second fundamental form $\thht_{AB}$ \eqref{adapth} which does not contain any antisymmetric combination of $\hsvd_B V_A$. Given \eqref{cos2}, the generic stress for the Cosserat fluid has antisymmetric contribution. The antisymmetric part of the Cauchy stress tensor for the Cosserat fluid is $\gamma_c (\p_{\mu} v_{\alpha} - \p_{\alpha} v_{\mu} - 2 \epsilon_{\mu \alpha \rho} w^{\rho})$. So the null fluid that we are trying to describe via \eqref{convective4} is a special case of the Cosserat fluid in the sense that the antisymmetric part of its own stress tensor will have no contribution from $\hsvd_B V_A - \hsvd_A V_B$. Having decided on this issue, we now compare $2 \gamma_c \p^{\mu}(- \epsilon_{\mu \alpha \rho} w^{\rho})$ and $\hsvd^B \Big[\frac{1}{8 \pi}\Big(-\fr q^d_{~B} q^c_{~A}T_{fcd}l^f\Big)\Big]$. Its clear that the spin $w^{\rho}$ of the Cosserat fluid particle is related to the torsion tensor on the null fluid side. This was perhaps anticipated since the origin of torsion is the spin angular momentum density in the matter action part. The extra material viscosity parameter $\g_c$ for the null fluid is hence identified as $\g_c = 1/(16 \pi)$. This establishes the connection that $\epsilon_{\mu \alpha \rho} w^{\rho}$ is related to $(1/2) T_{fAB}l^f$. In the characterization of the two dimensional null fluid we have to take notice of the fact that there exists the extra factor of $\hhw_B \hsvd_A V^B + {^{(2)}}T^B_{~CA}V^C \hhw_B$ along with the material derivative term in \eqref{convective1}. The analog of the first extra term in the case of Einstein gravity i.e $\Omega_B \mathcal{D}_A V^B$ was already present in the DNS equation.

In Einstein's gravity case it has been observed that the extra factors can be removed if the analysis is done in locally constructed inertial frame related to the adapted coordinates \cite{Padmanabhan:2010rp}. Then the Hajicek $1$-form equation exactly maps to NS dynamics. Having this instance, next we want to proceed our analysis in the local inertial frame. This would make the analogy more transparent.

\subsubsection{Analysis in a boosted local inertial frame constructed in adapted coordinates}\label{lif}
In this section we want to look at the evolution equation of $\hhw_A$ in the adapted coordinate system $x^i = (t, x^{\mu})$ i.e \eqref{convective1} from the perspective of a local inertial observer. However, before we go over to that particular analysis, we should comment on whether it is indeed possible to construct a local inertial frame (LIF) in the EC theory. The EC theory is provided with the non symmetric, however metric compatible Cartan connection $\bm{\hcvd}$. In fact, it has been pointed out that the only connection (with torsion) that is compatible with the Einstein equivalence principle is the Cartan connection \cite{Gogala1980TorsionAR, VonDerHeyde1975, Dey:2017fld}. The equivalence principle in the context of EC theory means the existence of a unique local frame where all the components of the Cartan connection coefficients vanish. If we insist on only holonomic or coordinate bases, then the only possible way for the connection coefficients (in the coordinate bases) to vanish is to demand that the torsion tensor vanishes. However, one can introduce a local Lorentz or orthonormal tetrad basis at the point $p$ in $\sptm$ where we want to construct our local inertial frame. The local orthonormal tetrad basis are required to be locally Minkowskian at the point $p$. With respect to this anholonomic tetrad bases, it has been explicitly shown \cite{VonDerHeyde1975, Dey:2017fld}, that it is indeed possible to set the Cartan connection coefficient (in the local orthonormal tetrad bases) to zero at the point $p$, without demanding that the torsion tensor also vanishes identically. We will hence not repeat the argument here. 

In order to study the dynamics of the null surface $\hech$ via \eqref{convective1} in the adapted coordinate system by constructing a LIF at a point $p$ on $\hech$, we need to revisit both the intrinsic and the extrinsic geometry of the null surface. Let us again reiterate the notion of the adapted coordinate system for clarity. The fact that we have foliated a neighborhood of the spacetime by a stack of spacelike $t(x^i) = \text{constant}$ hypersurfaces $\Sigma_t$, allows us to coordinatize the neighborhood by $x^i = (t, x^{\mu}) = (t, x^1 ,x^2 , x^3)$. The notion of the coordinate system being adapted or stationary w.r.t to the null surface basically means that the location of $\hech$ is fixed by the spatial coordinates by some scalar function say $f(x^1, x^2, x^3) = 0$ \cite{Gourgoulhon:2005ng}. One specific choice would be to set $f(x^1, x^2, x^3) = x^1 = 0$. We will henceforth work with this choice where $x^1 = 0$ defines the transverse two dimensional cross section $S_t$ on $\Sigma_t$. With this, the coordinates on $S_t$ surface are $x^{A} = (x^2, x^3)$ and the coordinate basis vectors of $T_{p}(S_t)$ are $\bm{e}_A = \bm{\p}_A = (\bm{e}_2, \bm{e}_3) = (\bm{\p}_2, \bm{\p}_3)$. The coordinate time evolution vector $\bm{t} = \bm{\p}_t$ connects same spatial points along neighboring $\Sigma_t$ hypersurfaces. Hence the coordinates defined on the null surface are $x^{\tilde{\mu}} = (t, x^A)$ with the coordinate basis vectors on $T_p(\hech)$ being $\bm{e}_{\tilde{\mu}} = (\bm{t}, \bm{e}_A) = (\bm{\p}_t, \bm{\p}_A)$. On the null surface $\hech$, we have $\bm{l} \overset{\hech}{=} \bm{t} + \bm{V}$, where $\bm{V}$ $\in$ $T_p(S_t)$. Hence w.r.t the coordinates on $\hech$ we have $l^{\tmu} \overset{\hech}{=} (1, V^A)$. It can be shown that the metric of the null surface $\hech$ w.r.t the coordinate system $x^{\tmu} = (t, x^A)$ is given by \cite{Gourgoulhon:2005ng, Padmanabhan:2010rp},
\begin{align}
ds^2_{|\hech} = q_{\tmu \tnu} dx^{\tmu} dx^{\tnu} = q_{AB}(dx^A - V^A dt)(dx^B - V^B dt) ~.
\label{indhech}
\end{align}
Here $\bm{q}$ represents the metric induced on $\hech$ from the metric $\bm{g}$ of the spacetime $\sptm$. $\bm{q}$ represents the first fundamental form of $\hech$ and provides a notion of the intrinsic geometry of $\hech$. From \eqref{indhech}, it can be easily verified that $l_{\tmu} = q_{\tmu \tnu}l^{\tnu} = (l_t, \{l_A\}) \oeq (0,0, 0)$. The null generator $\bm{l}$ belongs the tangent space $T_p(\hech)$ of $\hech$ and hence can be lowered w.r.t the induced metric $\bm{q}$. 

The Weingarten map $^{\hech}\chi^i_{~j}$ of $\hech$ represents the notion of the extrinsic curvature of $\hech$ \cite{Gourgoulhon:2005ng, Dey:2022qqp}. The Weingarten map was defined for any vector field $\bm{v} \in T_{p}(\hech)$ as,
\begin{align}
^{\hech}\chi^i_{~j} v^j = \hcvd_{\bm{v}}l^i ~.
\label{weinh}
\end{align}
Since the vector field $\bm{v}$ is arbitrary, we have as a result, $^{\hech} \chi^i_{~j} = \hcvd_j l^i$. With respect to the coordinates on $\hech$, we hence have, $^{\hech}\chi^{\tal}_{~\tbe} = \hcvd_{\tbe}l^{\tal}$. The second fundamental form $^{\hech}\bm{\Theta}$ restricted to $\hech$ is defined for arbitrary vectors $(\bm{u}, \bm{v}) \in T_{p}(\hech) \times T_{p}(\hech)$ in the following way \cite{Gourgoulhon:2005ng, Dey:2022qqp},
\begin{align}
^{\hech}\Theta_{ij} u^i v^j \equiv u_i ^{\hech}\chi^i_{~j} v^j ~.
\label{sech}
\end{align}
Let us make the choice of the vectors $\bm{u}$ and $\bm{v}$ to be the coordinate basis vectors $\bm{e}_{\tmu}$ and $\bm{e}_{\tnu}$ respectively. Then, we have,
\begin{align}
&^{\hech}\Theta_{ij} (\bm{e}_{\tmu})^i (\bm{e}_{\tnu})^j = (\bm{e}_{\tmu})_i~ ^{\hech}\chi^i_{~j} (\bm{e}_{\tnu})^j = (\bm{e}_{\tmu})_i (\hcvd_j l^i) (\bm{e}_{\tnu})^j = (\bm{e}_{\tmu})_i (\hcvd_j l^i) \d^j_{~\tnu} ~; \nonumber \\
&^{\hech}\Theta_{ij} \d^i_{\tmu} \d^j_{\tnu} = (\bm{e}_{\tmu})_i (\hcvd_{\tnu} l^i) = - l_i \hcvd_{\tnu} (\bm{e}_{\tmu})^i = -l_i (\hcvd_{\tnu}\d^i_{~\tmu}) = - l_i \hgm^i_{~\tnu j} \d^j_{~\tmu} ~; \nonumber \\
& ^{\hech}\Theta_{\tmu \tnu} = -l_i \hgm^i_{~\tnu \tmu} ~.
\label{sech1}
\end{align}
So we see that the second fundamental form restricted to the null surface $\hech$ is proportional to the connection coefficients of the RC spacetime. The non symmetric nature of $^{\hech}{\bm{\Theta}}$ is also evident from the fact that the connection coefficients are not symmetric. In fact, restricted to the transverse space $S_t$, we see that $^{\hech}\Theta_{A B} = - l_i \hgm^i_{BA}$. The above analysis leading to \eqref{sech1} can also be arrived from the relation \eqref{nablaalbexpansion}. Expressing \eqref{nablaalbexpansion} w.r.t the coordinates $x^{\tmu}$ on $\hech$, we have,
\begin{align}
&\hcvd_{\tnu} l_{\tmu} = \thht_{\tmu \tnu} + \hw_{\tnu} l_{\tmu} - l_{\tnu}(\hcvd_{\bm{k}}l_{\tmu}) \nonumber \\
& \implies \thht_{\tmu \tnu} = - \hgm^i_{~\tnu \tmu} l_i ~.
\end{align}
In the above, we have used the fact that $l_{\tmu} \oeq (0,0,0)$.

Next, we come to analysis of the rotation $1$-form, again restricted on $\hech$. In order to facilitate the form of the rotation $1$-form restricted to $\hech$, we would require a knowledge of the uniquely defined auxiliary null vector field $\bm{k}$. The auxiliary null vector field is not defined on the tangent space of $\hech$. Hence we would now require to choose a basis for $T_p(\mathcal{M})$. Let us define the outward pointing spacelike unit normal to the surface $S_t$ by $\bm{s}$ \cite{Gourgoulhon:2005ng}. Hence the basis vectors on $T_p(\Sigma_t)$ are $\bm{e}_{\mu} = (\bm{s}, \bm{e}_A)$. Extending the $\Sigma_t$ surfaces along the time evolution vector field, we have that the basis for $T_p(\mathcal{M})$ is $\bm{e}_i = (\bm{t}, \bm{s}, \bm{e}_A)$. The covector $k_i$ decomposed in such a basis is $k_i = (k_t, k_s, k_A)$. From the condition that $\bm{k} \cdot \bm{V} = 0$, we have $k_A = 0$. Similarly, the condition, $\bm{k} \cdot \bm{l} = -1$ leads us to have $k_t = -1$. Hence the expansion of the covector $k_i$ in the basis $(\bm{t}, \bm{s}, \bm{e}_A)$ is $k_i = (-1, k_s, 0)$. From the working definition of the rotation $1$-form \eqref{omega1defn}, we consider the projection of $\hhw_i$ onto the null surface $\hech$. The projection tensor onto $\hech$ is given by $\Pi^i_{~j} = \d^i_{~j} + k^i l_j$ \cite{Gourgoulhon:2005ng}. The projected part of the rotation $1$-form onto $\hech$ is defined by,
\begin{align}
^{\hech}\hw_i \equiv \Pi^j_{~i} \hw_j = - \Pi^j_{~i}(k_m \hcvd_j l^m)  = -k_m \hcvd_i l^m + l_i k_m k^j \hcvd_j l^m ~. 
\label{roth}
\end{align}
In the above, we have used the fact that $\Pi^j_{~i}l_j = 0$. We now specifically look at the components of $^{\hech}\hw_i$ in the coordinate basis of $T_p(\hech)$.
\begin{align}
^{\hech}\hw_{\tal} &= -k_m (\hcvd_{\tal}l^m) + l_{\tal} k_m (k^j \hcvd_j l^m) = -k_m (\hcvd_{\tal}l^m) \nonumber \\
& \oeq -k_0 \hcvd_{\tal}l^0 - k_1 (\hcvd_{\tal}l^1) -k_A (\hcvd_{\tal}l^A) \oeq \hcvd_{\tal}l^0 - k_1 (\hcvd_{\tal}l^1) \nonumber \\
^{\hech}\hw_{\tal} &\oeq \hgm^0_{~\tal j}l^j - \hgm^1_{~\tal j}l^j k_1 ~.
\label{roth1}
\end{align}
The projection of the rotation $1$-form onto the $2$-surface $S_t$ is the Hajicek $1$-form $\hhw_A$. Its clear that $^{\hech}\hw_{\tal} = (^{\hech}\hw_0 , {^{\hech}}\hw_A) = (^{\hech}\hw_0 , \hhw_A)$. This allows us to identify,
\begin{align}
\hhw_A = {^{\hech}}\hw_A \oeq \hgm^0_{A j}l^j - \hgm^1_{A j}l^j k_1 ~.
\label{roth2}
\end{align}
So, again w.r.t the coordinate basis established on $\hech$, we find that the rotation $1$-form (restricted to $\hech$) is  proportional to the connection coefficients. Similar analysis can also be brought through a tetrad basis $(\bm{n}, \bm{s}, \bm{e}_A)$ of $T_p(\mathcal{M})$, where $\bm{n}$ denotes the timelike unit normal to the $t(x^i) = \text{constant}$ surface. It can be shown that the Hajicek $1$-form w.r.t the above mentioned basis is associated to the tetrad connection coefficients \cite{Gourgoulhon:2005ch} as $\hhw_A = \hgm^1_{~0A}$. 

Having done this analysis, let us now look at the evolution equation for $\hhw_A$ i.e \eqref{convective4} in the adapted (or stationary) coordinates $x^{i} = (t, x^{\mu})$ w.r.t $\hech$. We propose to consider this expression around the given event point $p \in \hech$ in a LIF. In the LIF, the connection coefficients will vanish, but not their derivatives. We will be working with a Lorentz boosted inertial frame given by the fact that the metric coefficients are constant. The metric in this Lorentz boosted LIF is diagonal in structure with $V^A \neq 0$. The physical interpretation of such a boosted inertial frame has been explained in detail in \cite{Padmanabhan:2010rp}. Under the consideration of such a boosted LIF, all the terms that are proportional to the connection coefficients vanish. Hence $\hhw_A$ and $\thht_{BA}$ vanish in the LIF, however not their derivatives. Let us remember the fact that \eqref{convective4} has been derived under the geodesic constraint. Under this assumption, the trace of the second fundamental form $\thht_{BA}$ which is $(\thdl - T_b l^b )$, the shear tensor $\stsig_{BA}$ and the antisymmetric part $ \thht_{[BA]} = \fr q^c_{~A}q^d_{~B}T_{fdc}l^f = \fr [T_{0BA} + {^{2}}T_{DBA}V^D]$ all vanish in the boosted LIF. Since, we are working under the geodesic constraint, the second fundamental form is completely a spatial bilinear (given by the fact that $\thht_{ab}l^a = 0, \thht_{ab}l^b = 0, \thht_{ab}k^a = 0$ and $\thht_{ab}k^b = 0$). Let us look at the term $- \frac{1}{8 \pi}\Big(\thht_{ba}\hp^b q^a_{~A} -T_{iba} \thht^{bi} q^a_{A}\Big)$ in the external force density term \eqref{cos3}. The term within the parentheses (in the geodesic constraint) evaluates to $\Big(\thht_{CB}\hp^C q^B_{~A} - {^{2}T}_{CBD} \thht^{BC}q^D_{~A}\Big)$ which naturally vanishes in the boosted LIF. Finally, under these considerations, let us consider the evolution equation \eqref{convective4}, which now becomes (remembering that in the adapted coordinates $q^a_{A} = \d^a_{~B}$),
\begin{align}
&\p_t \Big(\frac{-\hhw_{A}}{8 \pi}\Big) + V^B \p_B \Big(\frac{-\hhw_A}{8 \pi}\Big)  =  -\d^b_{~A}\mt_{ab}l^a + \frac{1}{16 \pi} (\p_c +T_c) \Big(-S^c_{~ab}+ S_{ab}^{~~c} + S_{ba}^{~~c}\Big)l^a \d^b_{~A} \nonumber \\
& -\p_A \Big(\frac{\kappa}{8 \pi}\Big)+ \p^B \Big[\frac{1}{ 8 \pi} \Big(\stsig_{BA} - \fr \d_{AB}(\thdl - T_b l^b)\Big)\Big] + \p^B \Big[\frac{1}{8 \pi}\Big(-\fr \d^d_{~B} \d^c_{~A}T_{fcd}l^f\Big)\Big]~.
\label{convective3}
\end{align}
Even though the term $ \hhw_A (\thdl - T_b l^b)$ is zero in the LIF, it can be formally added so that its analogical structure with the Cosserat generalization to the NS fluid equation becomes evident. Now, in the boosted LIF, the equation \eqref{convective3} should be compared with the Cosserat fluid equation \eqref{cos2}. Here, we have the exact material derivative of the momentum density $\pi_A = - \hhw_A/8 \pi$. The null fluid pressure is $\kappa/8 \pi$ while the shear and extra material viscosity coefficients are $1/16 \pi$. The bulk viscosity coefficient is $- 1/16 \pi$. The external force density is $f^{\text{ext}}_{A} \equiv -\mt_{ab} \d^b_{~A}l^a + \frac{1}{16 \pi} (\p_c +T_c) \Big(-S^c_{~ab}+ S_{ab}^{~~c} + S_{ba}^{~~c}\Big)l^a \d^b_{~A}$ which quite naturally  arises from the term $-1/ 8 \pi \hg_{ab}l^a q^b_{~A}$, once the ECKS field equations are applied in the LIF.
This completes our analysis of the evolution equation \eqref{convective1} through an adapted coordinate system w.r.t $\hech$ about the point $p \in \hech$, where a boosted LIF has been constructed.

 A few comments are in order. Let us discuss again about the stress tensor of the Cosserat null fluid that we have been considering. The analogy was complete under the consideration that the stress tensor $\mathcal{S}_{AB}$ for our two dimensional null fluid living on $\hech$ in the EC theory has the form $\mathcal{S}_{BA} = 2 \eta \stsig_{BA} + \xi \d_{BA} (\thdl - T_b l^b) - 2 \gamma_c(1/2) T_{fAB}l^f$. It is obviously necessary that the trace free shear tensor is built from the derivative of the velocity field $V_A$ i.e we must have,
 \begin{align}
 \hsvd^B \stsig_{BA} = \fr \hsvd^B \Big[ (\hsvd_B V_B + \hsvd_A V_B) - \fr q_{AB}(\hsvd_C V^C) \Big] ~.
 \end{align}
 However, as evident from \eqref{adapth}, that would be possible only if $\p_t q_{AB} - 2 K_{0 (AB)} = \p_t q_{AB} - K_{0AB} - K_{0BA}$ was equal to zero. Now, it might be that in some choice of the adapted coordinate system, we might be able to set $\p_t q_{AB} = K_{0AB}+ K_{0AB}$. However, the interpretation of $\p_{t}q_{AB}$ in terms of fluid variables is under debate. This issue was also present in the DNS case \cite{Padmanabhan:2010rp}.
 In terms of the analogy, we pointed out that the momentum density of our null fluid in the EC theory is proportional to the Hajicek $1$-form $\hhw_a$ which a kinematical quantity. Its physical interpretation would be clearer if it was shown that the momentum density $\pi_A$ is indeed proportional to the velocity of the null fluid $V_A$. For the case of Einstein gravity, via the use of a particular adapted coordinate system w.r.t $\hech$, it was indeed shown that the Hajicek $1$-form was proportional to $V_A$ \cite{Kolekar:2011gw}. The same logic applies over here except that in the case of EC theory, we would have extra terms involving the two dimensional torsion tensor on $S_t$. We notice, that for our null fluid, both the shear viscosity and the extra material viscosity coefficients are positive. However, the bulk viscosity coefficient is negative. This feature was also present for the case of Einstein gravity. This is in contrast to the real world scenario where viscosity coefficients are positive. In the case of hydrodynamics, a negative bulk viscosity would lead to local entropy decrease with time. Reasons for the negative sign in the bulk viscosity has been attributed to the teleological nature of the horizon \cite{Gourgoulhon:2008pu, Bhattacharya:2015qkt, Bhattacharya:2017mrg, Cropp:2016ajh}. Finally, while considering the antisymmetric part of the stress tensor, we made the analogy that the quantity $\epsilon_{\mu \alpha \rho} w^{\rho}$ \eqref{cos2} is related to $(1/2 )T_{fAB}l^f$ for our null fluid \eqref{convective4}. One could have the viewpoint, that it might be better to relate $T_{fAB}$ to the intrinsic spin angular momentum tensor for the EC theory via the use of $T_{abc} = S_{abc} + (1/2) (g_{ab}S_c - g_{ac}S_b)$ and $S_{abc} = 8 \pi \tau_{abc}$, where $S_b = g^{ac}S_{abc}$. In that sense, perhaps the connection between the spin $w^{\mu}$ of the Cosserat fluid particle and the intrinsic spin density of our null fluid might have been more evident. This would have naturally resulted in a different material viscosity parameter $\gamma_c$. However, in dealing, with this analysis, we are of the viewpoint, that different kinematical quantities of $\hech$ are provided fluid interpretation (like $-\hhw_A/ 8 \pi$ as momentum density, $\kappa/8 \pi$ as the pressure etc). Similarly, here we adopt the geometrical viewpoint of torsion rather than that of a dynamical field. Such issues regarding whether torsion is to be interpreted as a geometrical or dynamical field has been explored in \cite{Dey:2017fld, Dey:2022qqp}.
 
\subsubsection{Covariant generalization of Cosserat fluid}
Finally, let us note that we have made a formal analogy between \eqref{convective1} and the Cosserat generalization to the NS fluid equation \eqref{cos2}. However instead, we could have taken the viewpoint that the natural generalization of the material derivative (in real world fluids) to the case of (hypersurface) null fluids on a genuinely curved background is the notion of the Lie derivative. Under that viewpoint, all the extra terms i.e $\hhw_B \hsvd_A V^B + {^{(2)}}T^B_{~CA}V^C \hhw_B$ can be effectively incorporated in the Lie derivative term as evident from \eqref{adapted1}. As a result, we have from \eqref{adapted1} and \eqref{convective1}, 
\begin{align}
&q^a_{~A} \lil \hhw_a + \hhw_A (\thdl - T_b l^b) \nonumber \\
& = 8 \pi q^b_{~A}\mt_{ab}l^a - \fr (\hcvd_c +T_c) \Big(-S^c_{~ab}+ S_{ab}^{~~c} + S_{ba}^{~~c}\Big)l^a q^b_{~A}+ \thht_{ba}\hp^b q^a_{~A} - T_{iba} \thht^{bi} q^a_{A}  \nonumber \\
&+ \hsvd_A \kappa- \hsvd_B \stsig^B_{~A} - \hsvd^B \Big(\fr q^c_{~A} q^d_{~B}T_{fdc}l^f\Big) + \fr \hsvd_A (\thdl - T_b l^b)~.
\label{convective2}
\end{align} 
The above dynamical equation for our null fluid can then be compared with the Cosserat generalization of the NS equation \eqref{cos2} with the identification that the material derivative has now been replaced with the Lie derivative term (for the null fluid).
\section{The tidal force equation in the RC spacetime $\sptm$}\label{tidalx}
We present the derivation of the tidal force equation in the RC spacetime for the sake of completeness. We will see that the tidal force equation is related to the evolution of the symmetric traceless shear tensor $\shdl$ along the null generators $\bm{l}$. As again, we begin with the Ricci identity,
\begin{align}
l^k\Big[\hcvd_k, \hcvd_j\Big]l_i = l^k\Big(-\hr^m_{~~ikj}l_m - T^m_{~~kj}\hcvd_m l_i\Big) \nonumber \\
\implies l^k\hcvd_k \hcvd_j l_i = l^k \hcvd_j (\hcvd_k l_i) - \hr_{mikj}l^m l^k - T^m_{~~kj}l^k(\hcvd_m l_i) ~. 
\label{tidal1}
\end{align}
Our analysis follows \cite{Gourgoulhon:2005ng}. 
With respect to \eqref{hphi}, we have the expansion for the covariant derivative of the null generators as,
\begin{equation}
\hcvd_a l_b = \hphi_{ba} + \hw_a l_b - l_a (k^i \hcvd_i l_b) - k_a q^c_{~b} \ti_c~.
\label{hphi1}
\end{equation}
 This allows us to have,
\begin{align}
&\qq l^r \hcvd_r \Big(\shdf_{ij} - \tit_{ij}\Big) + \shdf_{m}^{~~i} q_n^{~j}T_{irj}l^r - \tit_m^{~~i}q_n^{~j}T_{irj}l^r \nonumber \\
& + \fr \thdl \Big(2 \shdf_{mn} - \tit_{mn}- \qq T_{jri}l^r + \fr q_{mn}(\qtl) + \tit_{nm}\Big) \nonumber \\
& - \shdf_{mi} \tit^i_{~n} - \shdf_{ni} \tit_m^{~i} + q_{mn}(\shdf_{ij}\tit^{ij}) + \tit_{mi}\tit^i_{~n} - \fr q_{mn}(\tit^{ij}\tit_{ji}) \nonumber \\
& - (\kappa - \kt) \Big(\fr q_{mn}(\qtl) + \shdf_{mn} - \tit_{mn}\Big) + \fr q_{mn}\Big(l^k \hcvd_k (\qtl)\Big) \nonumber \\
& - \fr q_{mn} \Big(\shdf^{ij}T_{jri}l^r - \tit^{ij}T_{jri}l^r\Big) + q^r_{~m} \hhw_n \ti_r - \fr q_{mn}(q^{ij}\hhw_i \ti_j) - \hsvd_n (q^r_{~m}\ti_r) + \fr q_{mn} \hsvd_i (q^{ij}\ti_j) \nonumber \\
&+ q^b_{~m} q^d_{~n}\Big((\hcvd_c K_{adb} - \hcvd_d K_{acb}) + T^i_{~~cd}K_{aib} + (K^i_{~~cb}K_{adi} - K^i_{~~db}K_{aci})\Big)l^a l^c \nonumber \\
& - \fr q_{mn}\Big(\hcvd_i K^i_{~~ca} + \hcvd_c T_a + T^i_{~~jc}K^j_{~~ia} + K^i_{~~ja}K^j_{~~ci}+ T_i K^i_{~~ca}\Big)l^a l^c =  -  q^b_{~m} q^d_{~n} C_{abcd}l^a l^c ~.
\label{tidalforce}
\end{align}
A detailed derivation of the above result has been shown in Appendix \ref{tidal}.
Even though we are considering an integrable null hypersurface $\hech$ generated by $\bm{l}$ in the RC spacetime $\sptm$, the null generators are themselves null geodesics in the sense that they satisfy $l^b \nabla_b l^a = \kappa l^a$. However, in this general setup, the null generators are not autoparallel along themselves. In the context of a congruence of geodesic null curves, the term $-  q^b_{~m} q^d_{~n} C_{abcd}l^a l^c$ (present on the R.H.S of \eqref{tidalforce}) is related to the geodesic deviation equation \cite{wald2010general, padmanabhan2010gravitation} between two null geodesics. The driving force behind the relative acceleration between two neighboring null geodesics is directly related to the term involving the Weyl tensor in the R.H.S of \eqref{tidalforce}. It is in this respect that in literature, the above Eq. \eqref{tidalforce} is called the tidal force equation \cite{PhysRevD.33.915}. Our analysis also allows us to arrive at a different form of the NRE (as seen in \eqref{thdlevol}) corresponding to the outgoing expansion scalar $\thdl$ as compared to the ones presented in \cite{Dey:2017fld, Dey:2022qqp}. Let us mention that the tidal force equation for a null congruence has also been derived in \cite{PhysRevD.104.084073} (see Eq. ($63$) of \cite{PhysRevD.104.084073}). However, the traceless shear tensor considered in \cite{PhysRevD.104.084073} is different from $\shdf_{mn}$ analyzed over here. In \cite{PhysRevD.104.084073}, the shear tensor considered is the traceless symmetric part of the projected deviation tensor \cite{Dey:2022qqp} i.e $\hb_{ab} = q_a^{~i}q_b^{~j}B_{ij}$, where $B_{ij} = \hcvd_j l_i + T_{iaj}l^a$. As evident, from its definition, the projected deviation tensor is not symmetric. Its decomposition involves a non trivial antisymmetric part. However, our analysis leading to \eqref{tidalforce} deals with the deformation rate tensor \eqref{deformationrate} which by definition is symmetric. It can be shown that the trace part of both the deformation rate and the projected deviation tensor are equivalent \cite{Dey:2022qqp}. However, owing to the difference in their definitions, the symmetric traceless part of these two projected tensors ($\hb_{ab}$ and $\hchi_{ab}$) are different. 
\section{Discussions and comments}\label{conclusion}
As pointed out in \cite{Chakraborty:2015aja}, in the case of Einstein gravity, the structure of Einstein field equations near an arbitrary null surface can be understood w.r.t to the components of the vector field $G^a_{~b}l^b$ on the null surface. The components turn out to have very precise physical and thermodynamical interpretations. These insights have paved way to the understanding that the gravitational field equations (at least in the context of Einstein gravity) is perhaps emergent from underlying degrees of freedom associated with the gravitational field. Here, we have studied and tested the claim of emergence of the relevant gravitational field equations of a different theory of gravity. Our object of study has been the EC theory, which is the simplest of all possible gravitational theories under the umbrella of Poincaré gauge theory (PGT) of gravity \cite{aldrovandi2013teleparallel, Hehl:1976kj, de1986introduction, blagojevic2001gravitation, Blagojevic:2003cg}. The EC theory is a gravitational theory with non-propagating torsion i.e torsion cannot propagate outside the matter that sources the spin effects or the intrinsic angular momentum. The geometrical backdrop of the EC theory is the RC spacetime, in which our generic null hypersurface $\hech$ is constructed. As again, to understand the structure of the ECKS gravitational field equations about $\hech$, we consider the relevant projection components of $\hg^a_{~b}l^b$. They, as usual turn out to be $\hg_{ab}l^a l^b$, $\hg_{ab}k^a l^b$ and $\hg_{ab}q^a_{~c}l^b$. Naturally, the goal towards probing the emergent nature of the ECKS field equations would be to find out and analyze the physical underpinnings of these components. $\hg_{ab}l^a l^b$ is related to the dynamical evolution rule (the NRE) of the outgoing expansion scalar for the geodesic null congruence generating the integrable null hypersurface $\hech$. It was shown in \cite{Dey:2017fld}, that the application of NRE to a local thermodynamical constitutive relation about an approximate Rindler horizon established at any point $p$ $\in$ $\sptm$ gave emergence to the ECKS field equations. Similarly, it was shown by the authors of this paper in \cite{Dey:2022qqp}, that the component $\hg_{ab}k^a l^b$ is related to the dynamical evolution law of the ingoing expansion scalar of the null geodesic congruence forming $\hech$. Application of the process of virtual displacement of $\hech$ along the auxiliary null vector field allowed the interpretation of the ECKS field equations (via the component $\hg_{ab}k^a l^b$) as a structure analogous to the first law of thermodynamics. The physical interpretation of the component $\hg_{ab}q^a_{~c}l^b$ was missing in the literature.

In this work, we have shown that the transverse spatial component $\hg_{ab}q^a_{~c}l^b$ or rather $\hg_{ab}l^a q^b_{~c}$ is indeed related to the dynamical evolution rule of the Hajicek $1$-form for $\hech$. In the case of Einstein gravity, the correspondence of the dynamical evolution law for $\Omega^a$ in a set of coordinates adapted to $\hech$ yields a structure called the DNS equation. The DNS equation for the null fluid on $\hech$ is structurally very similar to the NS equation. The NS or the DNS equation is marked by the fact that its Cauchy stress tensor is symmetric by definition. This we expect, since the NS equation describes a fluid with no intrinsic angular momentum. This is turn, on the Einstein gravity side is guaranteed by the fact that there does not exist any contribution to the matter energy-momentum tensor due to the spin degrees of freedom in the microscopic domain. This, we surely cannot neglect in the EC theory. The source of torsion in the EC theory is precisely due to the intrinsic spin angular momentum tensor. The matter energy momentum sources the metric whereas torsion as  a geometric field is sourced by the spin angular momentum tensor. Hence, if it were indeed possible to interpret the evolution law for the Hajicek $1$-form $\hhw_a$ of $\hech$ in terms of a fluid/elastic model theory, then that particular theory must have a stress tensor that has a non-zero antisymmetric contribution. This we explicitly see from the term $\hsvd^b \Big(\fr q^c_{~t} q^d_{~b}T_{fdc}l^f\Big)$ in \eqref{lieomega1}. However, we reiterate that the evolution law had been derived under the geodesic constraint that forces the null generators of $\hech$ to be simultaneously auto-parallel and geodesic. 

We note that the presence of the antisymmetric contribution to the stress tensor in the case of EC theory hinted us to look at any real life scenario of fluids/elastic model systems with built in intrinsic angular momentum. One such hint came from \cite{vardoulakis2019cosserat, 2019EPJH...44...47S}. The Cosserat generalization to the NS equation describes a continuum fluid system in which the constituent material point bodies have translation as well as rotational degrees of freedom. Hence the Cosserat fluid does have an antisymmetric contribution to its stress tensor. Next, in order to furnish the analogy of the dynamical evolution equation of $\hhw_a$ with that of fluid dynamical equation of the Cosserat fluid, we adopted two ways. In the first approach, we expressed the evolution equation of $\hhw_a$ in terms of a coordinate system $(t,x^{\mu})$ adapted to the null hypersurface $\hech$ \eqref{convective1}. Making suitable identification of the external body force density \eqref{cos3}, we compared our resulting equation with the Cosserat fluid equation \eqref{cos2}. The momentum density of our null fluid on $\hech$ turned out to be $- \hhw_A/(8 \pi)$. The comparison allowed us to extract the viscosity coefficients with the shear and the extra material viscosity parameter $\gamma_c$ turning out to be positive. The bulk viscosity coefficient of our null fluid as in the Einstein case remains negative. We then performed the analysis in a boosted LIF which made the analogy of our dynamical evolution equation of $\hhw_A$ with the Cosserat fluid more evident. In the second approach, we made the identification that the material derivative of the Cosserat fluid (written in Cartesian coordinates in Galilean spacetime) should be replaced by the Lie derivative. This is on the basis that the Lie derivative of the momentum density should be the natural generalization of the material derivative in a genuinely curved background.

Having done this analysis, we should be cautious about the analogy brought in. There are certain issues regarding the fluid interpretation of the gravitational field equations which are in still under debate. Firstly, w.r.t \eqref{adapth}, consider the quantity, $\p_t q_{AB}$
. It is w.r.t a convenient choice of the adapted coordinates w.r.t $\hech$, can we set $\p_t q_{AB}$ to zero. However, the physical interpretation of this quantity in terms of fluid variables is still lacking. This was also the case for Einstein gravity. In addition, for the case of the ECKS theory, we understand, w.r.t to \eqref{adapth}, there is an extra contribution to the extended second fundamental form $\thht_{BA}$ written in the adapted set of coordinates $(t,x^{\mu})$ i.e $\fr(- 2 K_{0AB} - {^{(2)}}T_{DAB}V^D)$. The antisymmetric part of the second fundamental form i.e $ \thht_{[BA]} = 1/2(T_{0BA} + {^{(2)}}T_{DBA}V^D) = 1/2 (q^d_{~B} q^c_{~A}) T_{fdc}l^f$ should be related to spin $w^{\mu}$ of the Cosserat fluid particle \eqref{cos2}. However, the symmetric part of $\thht_{BA}$ barring aside the term $1/2(\hsvd_A V_B + \hsvd_B V_A)$ (analogous to the strain rate or classical deformation rate tensor $D_{\mu \nu}$ in \eqref{cos1}) i.e the term $\p_t q_{AB} -K_{0AB} - K_{0BA}$ does not have direct fluid interpretation. It is again only by a convenient choice of the adapted coordinates can we set this extra term to zero.
 Hence the correspondence of the dynamical evolution law for $\hhw_a$ with the Cosserat fluid is in no way watertight. Further scrutiny and insight into the analogy is desired.

 For a real world fluid, negative value of bulk viscosity coefficient would imply a dilation or contraction instability. This would translate to the fact, that the global null surface $\hech$ is unstable under external perturbations. The fact, that the bulk viscosity coefficient is negative is in agreement with the fact that a generic hypersurface has the tendency to continually contract to expand. For the case of event horizon, the expansion vanishes under the equilibrium condition attained at far future and hence it stabilizes \cite{Gourgoulhon:2008pu}. In any case, the event horizon is a global concept, that require full knowledge of the spacetime or complete future predictability of any Cauchy surface. There are local concepts of horizons which are bereft of this teleological property. It has been shown that the generalized DNS equation applied to future outer trapping horizon and dynamical horizon in $(\mathcal{M}, \bm{g}, \bm{\nabla})$ leads to a positive bulk viscosity coefficient \cite{Gourgoulhon:2008pu, Gourgoulhon:2005ch}. It is also desirable to see whether we  encounter the same positive bulk viscosity coefficient for such local horizons in the RC spacetime.

This ``completes'' the physical interpretation for the trifecta of the relevant projection components of the vector field $\hg^a_{~b}l^b$ in the EC theory. The fact that the ECKS field equations expressed w.r.t to a generic null surface $\hech$ has both thermodynamic (via $\hg_{ab}l^a l^b$ and $\hg_{ab}k^a l^b$) and fluid dynamic (via $\hg_{ab}q^a_{~c}l^b$) interpretation lends strength to the concept of the emergent paradigm of gravity. 

For the sake of completeness, we also computed the tidal force equation for a null geodesic congruence in the RC spacetime. Here, we did not employ the geodesic constraint and kept the analysis general. The only assumption that went into the tidal force equation was that the geodesic null congruence forms an integrable hypersurface, i.e the Frobenius identity is satisfied. 

 A very natural question that may arise is whether the same analysis would be possible in the language of tetrad variables. In the tetrad approach, the veilbein and  Lorentz spin connection are the dynamical variables in place of metric and  torsion. In the EC theory, the connection is metric compatible which translates to the spin connection being antisymmetric in the tetrad language. Even though we have not worked out explicitly the details, the analysis in terms of the tetrad variables should finally result in the same dynamical evolution law for the Hajicek $1$-form. In fact, it would be worthwhile to carry out the analysis in PGT, which naturally uses the language of tetrads. In that, the EC and teleparallel gravity theory are the most discussed ones. We hope to come back to this analysis in a future work.

 \section*{Acknowledgments}
 The research of one of the authors (B.R.Majhi) is supported by Science and Engineering Research Board (SERB), Department of Science $\&$ Technology (DST), Government of India, under the scheme Core Research Grant (File no. CRG/2020/000616).
 
 \appendix
\section*{Appendices}
\section{Derivation of the relation \eqref{evo1}}\label{evol1}
Let us begin with the first term within the parentheses in the L.H.S of \eqref{Ricciiden}. Using \eqref{relnnablaalbnchi} in this term, one finds,
\begin{align}
\hcvd_b (\hcvd_a l^b) &= \hcvd_b \Big[\hchi_a^{~b} + \hw_a l^b -l_a(k^i \hcvd_i l^b) - k_a q^{cb} \ti_c - q^c_{~a} q^{db}K_{fcd}l^f\Big] ~.
\label{manip1}
\end{align}
Then, we make use of \eqref{nablaakbexpansion} and \eqref{expansionscalervalue} in \eqref{manip1} for the spacetime covariant derivative of the ingoing auxiliary null vector field $k^a$ and the covariant divergence of the null generator $l^a$ respectively. Upon simplification, this leads to,
\begin{align}
\hcvd_b (\hcvd_a l^b) &= \hcvd_b \hchi^b_{~a} + l^b \hcvd_b \hw_a + \hw_a \Big(\thdl+ \kappa -T_il^i\Big) - \hchi_{ab}(k^j\hcvd_j l^b) + q^c_{~a} \ti_c (k_b(k^j \hcvd_j l^b)) \nonumber \\
& + q^d_{~a}q^c_{~b} K_{fcd}l^f(k^j \hcvd_j l^b) - \xih_a^{~c} \ti_c + q^d_{~a} q^{ci} \ti_i K_{fcd}k^f - \hcvd_b (q^c_{~a}q^{db}K_{fcd}l^f) \nonumber \\
& -l_a\Big(\hw_b(k^j \hcvd_j l^b) + \hcvd_b(k^j \hcvd_j l^b)\Big) + k_a \Big(\hhw^c \ti_c - \hcvd_b(q^{cb}\ti_c)\Big) ~.
\label{manip2}
\end{align}
Similarly, taking on the second term in the commutator bracket of L.H.S of \eqref{Ricciiden}, we have upon using \eqref{expansionscalervalue},
\begin{align}
\hcvd_a(\hcvd_b l^b) &= \hcvd_a (\thdl + \kappa) - (\hcvd_a T_i)l^i - T^i \hchi_{ia} - \hw_a (T^i l_i) + l_a \Big(T^i (k^j \hcvd_j l_i)\Big) + k_a (q^c_{~i}\ti_c T^i) \nonumber \\
& + q^c_{~a} q^d_{~b}T^b(K_{fcd}l^f) ~.
\label{manip3}
\end{align}
Using the above results we compute the quantity $\Big[\hcvd_b, \hcvd_a\Big]l^b $. We then project it on the two-surface $S_t$ and have the expression for the L.H.S of \eqref{projRicciiden},
\begin{align}
\Big[\hcvd_b, \hcvd_a\Big]l^b q^a_{~t} &= q^a_{~t} \hcvd_b \Big[\hchi^b_{~a} - q^c_{~a} q^{db}K_{fcd}l^f\Big] + q^{a}_{~t}(l^b \hcvd_b \hw_a) + \hhw_a \Big(\thdl + \kappa\Big) - \hchi_{tb} (k^j \hcvd_j l^b) \nonumber \\
& + q^c_{~t} \ti_c k_b(k^j \hcvd_j l^b) - \xih^c_{~t} \ti_c - \hsvd_t \Big(\thdl + \kappa\Big) + q^a_{~t}(\hcvd_a T_i)l^i + \hchi_{ti} T^i \nonumber \\
& + q^c_{~t} q^d_{~b} \Big[(K_{fdc}l^f)(k^j \hcvd_j l^b) + (K_{fdc}k^f) \ti^b - (K_{fcd}l^f)T^b\Big] ~.
\label{LHS1}
\end{align}

Let us now, focus on the term $q^a_{~t} \hcvd_b \Big[\hchi^b_{~a} - q^c_{~a} q^{db}K_{fcd}l^f\Big]$ in R.H.S of \eqref{LHS1} and try to manipulate it. 
We see then,
\begin{align}
&q^a_{~t} \hcvd_b \Big[\hchi^b_{~a} - q^c_{~a} q^{db}K_{fcd}l^f\Big] = q^a_{~t} \hcvd_b \hphi^b_{~a} = q^a_{t} \delta^j_{~b} \d^k_{~a} (\hcvd_j \hphi^b_{~k}) \nonumber \\
&= q^a_{~t}(q^j_{~b}- l^j k_b -k^j l_b)(q^k_{~a} - l^k k_a- k^k l_a) \hcvd_j \hphi^b_{~k} = q^j_{~b} q^a_{~t}q^k_{~a} \hcvd_j \hphi^b_{~k} - q^k_{~t}l^j k_b \hcvd_j \hphi^b_{~k} - q^k_{~t} k^j l_b \hcvd_j \hphi^b_{~k} \nonumber \\
& = \hsvd_b \hphi^b_{~t} + q^k_{~t} \hphi^b_{~k}(l^j \hcvd_j k_b) + q^k_{~t} \hphi^b_{~k}(k^j \hcvd_j l_b) = \hsvd_b \hphi^b_{~t} + q^k_{~t} \hphi^b_{~k}\Big(\hw_b - \hp_b\Big) + q^k_{~t} \hphi^b_{~k}(k^j \hcvd_j l_b) \nonumber \\
& = \hsvd_b \hphi^b_{~t} + \hphi^b_{~t}\Big(\hhw_b - \hp_b\Big) + \hchi_{bt}(k^j \hcvd_j l^b) - q^c_{~t} q^d_{~b}(K_{fcd}l^f)(k^j \hcvd_jl^b) ~.
\label{manip5}
\end{align}
In the above, we define the spatial covector $\hp_a$ as $\hp_a \equiv T_{bcd} k^b l^c q^d_{~a}$. In arriving at \eqref{manip5}, we have used orthogonality relations of the spatial tensor $\hphi^a_{~b}$ (w.r.t $\bm{l}$ and $\bm{k}$) and \eqref{rotn1forndefn2} in the third line. Similarly, we have used \eqref{Omegadefn1} in the final fourth line.
Next, we focus on the term $q^a_{~t}(l^b \hcvd_b \hw_a)$ of \eqref{LHS1}. Using \eqref{rotnhajireln}, we have,
\begin{align}
q^a_{~t}(l^b \hcvd_b \hw_a) = q^a_{~t} l^b \hcvd_b \Big(\hhw_a - \kappa k_a + (\bm{k}\cdot  \mathbbb{T})k_a\Big) = q^a_{~t}(l^b \hcvd_b \hhw_a) - (\kappa - \kt ) q^b_{~t}(\hhw_b -\hp_b) ~.
\label{manip6}
\end{align}
In the above, we have as usual used \eqref{rotn1forndefn2} and \eqref{Omegadefn1}. 
Note that using \eqref{relnnablaalbnchi} for the expansion of the spacetime covariant derivative of the null generator in terms of the deformation rate tensor, it can quite easily be shown that,
\begin{align}
q^a_{~t}(l^b \hcvd_b \hhw_a) = q^a_{~t} \lil \hhw_a - \hchi_{tb} \hhw^b + q^c_{~t} \hhw^d K_{fcd}l^f - q^a_{~t}T_{bca} \hhw^b l^c ~. 
\label{manip7}
\end{align}
Using this in \eqref{manip6}, we finally obtain,
\begin{align}
q^a_{~t}(l^b \hcvd_b \hw_a) =  q^a_{~t} \lil \hhw_a - \hchi_{tb} \hhw^b - (\kappa - \kt) (\hhw_t - \hp_t) + q^a_{~t} \hhw^b l^c(K_{cab} - T_{bca}) ~.
\label{manip8}
\end{align}
Upon using \eqref{manip8} and \eqref{manip5} in \eqref{LHS1}, we obtain, after a bit simplification,
\begin{align}
&\Big[\hcvd_b, \hcvd_a\Big]l^b q^a_{~t} = \hsvd_b \hphi^b_{~t} + \hphi^b_{~t} (\hhw_b - \hp_b) + q^a_{~t} \lil \hhw_a + \hchi_{ti} (T^i - \hhw^i)\nonumber \\
& - \xih_{ti} \ti^i - (\kappa - \kt) (\hhw_t - \hp_t) + q^a_{~t} \hhw^b l^c (K_{cab} - T_{bca}) + \hhw_t (\thdl + \kappa) + q^c_{~t} \ti_c k_b (k^j \hcvd_j l^b)\nonumber \\
& - \hsvd_t(\thdl + \kappa) + q^a_{~t} (\hcvd_a T_i)l^i  + q^c_{~t} q^d_{~b} \Big[(T_{fdc}l^f)(k^j \hcvd_j l^b) + (K_{fdc}k^f) \ti^b - (K_{fcd}l^f)T^b\Big] ~.
\label{LHS2}
\end{align}
We manipulate further a few terms on the R.H.S of \eqref{LHS2}. We note that upon using the definition \eqref{hphi}, we have,
\begin{align}
\hphi_{bt} (\hhw^b - \hp^b) + \hchi_{bt} (T^b - \hhw^b) = \hchi_{bt}(T^b - \hp^b) - q^c_{~t} \hhw^d (K_{fcd}l^f) + q^c_{~t} \hp^d (K_{fcd}l^f) ~.
\label{manip9}
\end{align}
Next, we manipulate the term $q^a_{~t}(\hcvd_a T_b)l^b$ in the R.H.S of \eqref{LHS2} via using \eqref{relnnablaalbnchi}:
\begin{align}
q^a_{~t}(\hcvd_a T_b)l^b &= q^a_{~t}\Big(\hcvd_a (T_b l^b) - T^b (\hcvd_a l_b)\Big) \nonumber \\
& = \hsvd_t (T_b l^b) - \hchi_{ta}T^a - \hhw_t (T_a l^a) + q^c_{~t} q^d_{~b}T^b (K_{fcd}l^f) ~.
\label{manip10}
\end{align}
Finally using the relations \eqref{manip9} and \eqref{manip10} in \eqref{LHS2}, we obtain after simplification,
\begin{align}
\Big[\hcvd_b, \hcvd_a\Big]l^b q^a_{~t} &= \hsvd_b \hphi^b_{~t} + q^a_{~t} \lil \hhw_a  - \hchi_{bt} \hp^b + q^c_{~t} \hp^d (K_{fcd}l^f) - \xih_{ti} \ti^i   \nonumber \\
& - (\kappa - \kt) \hhw_t + (\kappa - \kt) \hp_t - q^a_{~t} \hhw^b l^c T_{bca} + \hhw_{t}\Big(\thdl+ \kappa\Big) + q^c_{~t} \ti_c k_b (k^j \hcvd_j l^b) \nonumber \\
& - \hsvd_t\Big(\thdl + \kappa - T_b l^b\Big) - \hhw_t (T_a l^a)  + q^c_{~t} q^d_{~b} (T_{fdc}l^f)(k^j \hcvd_j l^b) + q^c_{~t}q^d_{~b} (K_{fdc}k^f) \ti^b 
\label{LHS3}
\end{align}

Now, we bring our focus to the R.H.S of \eqref{projRicciiden}. As usual, we use \eqref{relnnablaalbnchi} for the expansion of the covariant derivative of the null generator. This yields,
\begin{align}
\hr_{ab}l^a q^b_{~t} - T^i_{~ba}(\hcvd_i l^b)q^a_{~t} &= \hr_{ab}l^a q^b_{~t} - T^i_{~ba} \hchi^b_{~i} q^a_{~t} - (T_{iba} \hw^i l^b)q^a_{~t} + (T_{iba}l^i)(k^j \hcvd_j l^b)q^a_{~t} \nonumber \\
& + T_{iba} k^i q^{cb}\ti_c q^a_{~t} + T^i_{~ba}q^c_{~i} q^{db} (K_{fcd}l^f)q^a_{~t} ~.
\label{manip11}
\end{align}
We now express the term $(T_{iba}l^i)(k^j \hcvd_j l^b)q^a_{~t}$ in the R.H.S of \eqref{manip11} in a different way:
\begin{align}
(T_{iba}l^i)(k^j \hcvd_j l^b)q^a_{~t} &= q^c_{t} \d^d_{~b} (T_{fdc}l^f) (k^j \hcvd_j l^b) = q^c_{~t}(q^d_{~b} - k^d l_b - l^d k_b)(T_{fdc}l^f)(k^j \hcvd_j l^b) \nonumber \\
& = q^c_{~t} q^d_{~b}(T_{fdc}l^f)(k^j \hcvd_j l^b) + q^c_{~t} k_b \ti_c (k^j \hcvd_j l^b) ~.
\label{manip12}
\end{align}
Upon using \eqref{manip12} in \eqref{manip11} and using \eqref{rotnhajireln}, we obtain,
\begin{align}
\hr_{ab}l^a q^b_{~t} - T^i_{~ba}(\hcvd_i l^b)q^a_{~t} &=\hr_{ab}l^a q^b_{~t} - T_{iba} \hchi^{bi}q^a_{~t} - T_{iba} \hhw^i l^b q^a_{~t} + (\kappa - \kt) \hp_t \nonumber \\
& + q^c_{~t} q^d_{~b}(T_{fdc}l^f)(k^j \hcvd_j l^b) + q^c_{~t} k_b \ti_c (k^j \hcvd_j l^b) + T_{iba} k^i q^{cb} \ti_c q^a_{~t} \nonumber \\
& + q^a_{~t} q^{ci} q^{db} T_{iba} K_{fcd}l^f ~.
\label{RHS}
\end{align}
Now, all we got to do is to equate \eqref{LHS3} and \eqref{RHS} via \eqref{projRicciiden} and simplify. To this effect, we obtain \eqref{evo1}.
\section{Derivation of the relations \eqref{adapchi} and \eqref{adapphi}} \label{adapx}
  We will now expand the deformation rate tensor $\hchi_{ij} = \fr q_i^{~m} q_j^{~m} \lil q_{mn}$ w.r.t to the coordinate system $(t, x^{\mu})$ generated by the foliation of $\sptm$ by the stack of spacelike hypersurfaces $\Sigma_t$. Our analysis follows \cite{Gourgoulhon:2005ng}. With respect to the $(3+1)$ foliation, we have that $\bm{l} = \bm{t} + \bm{V} + (N-b)\bm{s}$, where $N$ is the lapse function and $\bm{s}$ is the outward pointing spacelike unit normal to transverse cross section $S_t$ on $\Sigma_t$. The orthogonal decomposition of the time evolution vector field $\bm{t}$ is given by $\bm{t} = N \bm{n} + \bm{\b}$. The spatial shift vector $\bm{\b}$ can again be provided an orthogonal decomposition on $\Sigma_t$ via $\bm{\b} = b \bm{s} - \bm{V}$, where $\bm{V}$ is a vector field established on the tangent space of $S_t$. As a consequence, we have,
  \begin{align}
  \hchi_{ij} = \fr q_i^{~r} q_j^{~s} \Big[\pounds_{\bm{t}} q_{rs} + \pounds_{\bm{V}} q_{rs} + \pounds_{(N-b)\bm{s}}q_{rs}\Big] ~.
  \label{manip35}
  \end{align}
  Expanding the term $\pounds_{\bm{V}} q_{rs}$ in terms of the spacetime covariant derivative, we have,
\begin{align}
\pounds_{\bm{V}} q_{rs} = V^i \hcvd_i q_{rs} + q_{ri} (\hcvd_s V^i) + q_{is}(\hcvd_r V^i) + T^k_{~ir} q_{ks}V^i + T^k_{~is}q_{rk}V^i ~.
\label{manip36}
\end{align}
Same expansion for the term $\pounds_{(N-b)\bm{s}}q_{rs}$ leads us to,
\begin{align}
\pounds_{(N-b)\bm{s}}q_{rs} &= (N-b)s^i \hcvd_i q_{rs} + q_{ri} \hcvd_s[(N-b) s^i] + q_{is}\hcvd_r [(N-b)s^i] + T^k_{~ir} q_{ks}(N-b)s^i\nonumber \\
& + T^k_{~is}q_{rk}(N-b)s^i ~.
\label{manip37}
\end{align}
Putting the relations \eqref{manip36} and \eqref{manip37} in \eqref{manip35}, and using the fact that $q_i^{~r}q_j^{~s}V^i \hcvd_i q_{rs} = 0$ and $(N-b)q_i^{~r}q_j^{~s}s^i \hcvd_i q_{rs} = 0$, we simplify the resulting expression for $\hchi_{ij}$. to have,
\begin{align}
\hchi_{ij} &= \fr \Big[q_i^{~r}q_j^{~s} \pounds_{\bm{t}}q_{rs} + \hsvd_j V_i + \hsvd_i V_j + {^{(2)}}T_{jli} V^l + {^{(2)}}T_{ilj}V^l \nonumber \\
& + (N-b)\Big(\hat{H}_{ij} + \hat{H}_{ji} + q_i^{~r}q_j^{~k}T_{klr}s^l + q_i^{~k}q_j^{~r}T_{klr}s^l\Big)\Big]~. \label{manip38}
\end{align}
In the above, we have used the following definitions. The spatial covariant derivative (compatible with the induced metric $\bm{q}$ of $S_t$) of any vector field $\bm{V}$ lying in the tangent space of $S_t$ is given by $\hsvd_i V_j \equiv q_i^{~r}q_j^{~k}(\hcvd_r V_k)$. The two dimensional torsion tensor ${^{(2)}}T^a_{~bc}$ as result of the induced connection $\bm{\hsvd}$ on the submanifold $(S_t, \bm{q})$ is given via the relation,
\begin{align}
{^{(2)}}T^a_{~bc} = q^a_{~m}q_b^{~d}q_c^{~f}T^m_{~~df} ~.\label{manip39}
\end{align}
The proof of \eqref{manip39} has been presented in \cite{Dey:2022qqp}. Similarly, the extrinsic curvature $H_{ij}$ of the $2$-surface $S_t$ viewed as embedded hypersurface in the $3$-dimensional spacelike surface $\Sigma_t$ is given by,
\begin{align}
H_{ij} = q_i^{~r}q_j^{~k}(\hcvd_r s_k) ~.
\end{align}
Now, for an adapted coordinate system w.r.t $\hech$, we have $b \overset{\hech}{=} N$ and $q^i_{A} = \d^i_{~A}$ \cite{Gourgoulhon:2005ng}. Thus on the null hypersurface, we have $\bm{l}  \overset{\hech}{=} \bm{t} + \bm{V}$. As a result, from \eqref{manip38}, we have,
\begin{align}
\hchi_{ij} \overset{\hech}{=} \fr \Big[q_i^{~r}q_j^{~s} \pounds_{\bm{t}}q_{rs} + \hsvd_j V_i + \hsvd_i V_j + \Big({^{(2)}}T_{jli}  + {^{(2)}}T_{ilj}\Big)V^l \Big]~.
\label{manip40}
\end{align}
Similarly, the spatial tensor $\tit_{ij}$ can be expressed in the following way on the null surface $\hech$,
\begin{align}
\tit_{ij} = q_i^{~s}q_j^{~r}K_{trs}l^t \overset{\hech}{=}q_i^{~s}q_j^{~r}K_{trs} (t^t + V^t) = (q_i^{~s}q_j^{~r}K_{mrs}t^m) + {^{2}}K_{tji}V^t ~.
\label{manip41}
\end{align}
Finally, the spatial tensor $\hphi_{ij}$, turns out to be,
\begin{align}
\hphi_{ij} = \hchi_{ij} - \tit_{ij}\overset{\hech}{=} \fr \Big[q_i^{~r}q_j^{~s} \pounds_{\bm{t}}q_{rs} + \hsvd_j V_i + \hsvd_i V_j + \Big({^{(2)}}T_{jli}  + {^{(2)}}T_{ilj}\Big)V^l  - 2 q_i^{~s}q_j^{~r}K_{mrs}t^m - 2( {^{2}}K_{lji})V^j\Big] ~.\label{manip42}
\end{align} 
Using the fact that for the adapted coordinate system $q^i_{~A} = \d^i_{~A}$, we have via \eqref{manip42},
\begin{align}
\hphi_{AB} \overset{\hech}{=} \fr \Big[\pounds_{\bm{t}}q_{AB} + \hsvd_A V_B + \hsvd_B V_A + \Big({^{2}}T_{ADB} + {^{2}}T_{BDA}\Big)V^D - 2 K_{0BA} - 2({^{2}}K_{DBA})V^D\Big] ~.\label{manip43}
\end{align}
Using the definition of contorsion tensor, we easily arrive at \eqref{adapphi}. Similar analysis on \eqref{manip40} leads us to \eqref{adapchi}.
\section{Derivation of \eqref{tidalforce}}\label{tidal}
Let us manipulate the first term in the R.H.S of \eqref{tidal1}, by repeated use of \eqref{relnnablaalbnchi}.
Upon using, \eqref{relnnablaalbnchi} on the term $l^m \hcvd_j (\hcvd_m l_i)$, we have, 
\begin{align}
&l^m \hcvd_j (\hcvd_m l_i) = l^m \hcvd_j (\hphi_{im}) + (l^m \hcvd_j \hw_m) l_i + (\kappa -\kt) (\hcvd_j l_i) - (l^m \hcvd_j k_m) q^r_{~i} \ti_r \nonumber \\
&+ \hcvd_j (q^r_{~i} \ti_r) - \hphi_{im}(\hcvd_j l^m) + (l^m \hcvd_j \hw_m)l_i + (\kappa - \kt)(\hcvd_j l_i) - l^m (\hcvd_j k_m)q^r_{~i} \ti_r + \hcvd_j (q^r_{~i}\ti_r) ~.
\label{manip16}
\end{align}
We now use the fact that the $(0,2)$ tensor $\hphi_{ij}$ is a completely spatial tensor and hence, $l^m \hcvd_j (\hphi_{im}) = - \hphi_i^{~m} (\hcvd_j l_m)$, upon which we again make use of \eqref{relnnablaalbnchi}. This leads us to,
\begin{align}
&l^m \hcvd_j (\hcvd_m l_i) = -\hphi_i^{~m} \hphi_{mj} + l_j \hphi_{im} (k^r\hcvd_r l^m) + k_j q^s_{~m}  \ti_s \hphi_i^{~m} - q^r_{~i} \ti_r \hw_j- l_j \Big(q^r_{~i} \ti_rk^m (k^s \hcvd_s l_m)\Big) \nonumber \\
& + l_i (l^m \hcvd_j \hw_m) + (\kappa - \kt) \hphi_{ij} + l_i (\kappa - \kt) \hw_j - l_j \Big[(\kappa - \kt) (k^s \hcvd_s l_i)\Big] \nonumber \\
& - k_j \Big[(\kappa - \kt)q^s_{~i}\ti_s\Big] + \hcvd_j (q^r_{~i}\ti_r) ~.
\label{manip17}
\end{align}
Let us look at the third term $-T^m_{~~kj}l^k(\hcvd_m l_i)$ in the R.H.S of \eqref{tidal1}. Again, upon use of the relation \eqref{relnnablaalbnchi}, and the symmetry property of the torsion tensor, we have,
\begin{align}
-T^m_{~~kj}l^k(\hcvd_m l_i) &= - \hphi_{im}T^m_{~~kj}l^k - \ti_j (k^s \hcvd_s l_i) + \hp_j q^r_{~i} \ti_r - (T_{mkj} \hw^m l^k) l_i \nonumber \\
& + l_j (T_{mks}k^m l^k k^s) q^r_{~i} \ti_r ~.
\label{manip18}
\end{align}
In principle, we would like to project the Eq. \eqref{tidal1} onto the $2$-surface $S_t$. As a consequence, we have,
\begin{align}
q^i_{~m} q^j_{~n} l^k \hcvd_j (\hcvd_k l_i) = - \hphi_{mt} \hphi^t_{~n} - \hhw_n (q^r_{~m} \ti_r) + (\kappa - \kt) \hphi_{mn} + q^i_{~m} q^j_{~n} \hcvd_j (q^r_{~i}\ti_r) ~.
\label{manip19}
\end{align}
Similarly, 
\begin{align}
-q^i_{~m} q^j_{~n} T^r_{~~kj}l^k(\hcvd_r l_i) = - \hphi_{mr} T^r_{~~kj}l^k q^j_{~~n} - (q^j_{~n} \ti_j) (q^i_{~m} k^s \hcvd_s l_i) + \hp_n q^r_{~m} \ti_r ~.
\label{manip20}
\end{align} 
Thus upon using \eqref{manip19} and \eqref{manip20}, we have, 
\begin{align}
&q^i_{~m} q^j_{~n}\Big(l^k \hcvd_j (\hcvd_k l_i) - \hr_{mikj}l^m l^k -T^r_{~~kj}l^k(\hcvd_r l_i) \Big) = - \hphi_{mt} \hphi^t_{~n} - \hhw_n (q^r_{~m} \ti_r) + (\kappa - \kt) \hphi_{mn}\nonumber \\
& + q^i_{~m} q^j_{~n} \hcvd_j (q^r_{~i}\ti_r) - q^i_{~m} q^j_{~n}\hr_{risj}l^r l^s - \hphi_{mr} T^r_{~~kj}l^k q^j_{~~n} - (q^j_{~n} \ti_j) (q^i_{~m} k^s \hcvd_s l_i) + \hp_n q^r_{~m} \ti_r~.
\label{manip21}
\end{align}
Let us focus back on the term in the L.H.S of \eqref{tidal1}. As usual, use of \eqref{relnnablaalbnchi} and \eqref{rotn1forndefn2} on the term $l^k \hcvd_k (\hcvd_j l_i)$ leads to,
\begin{align}
l^k \hcvd_k (\hcvd_j l_i) &= l^k \hcvd_k \hphi_{ij} + l_i (l^k \hcvd_k \hw_j) + \hw_j (\kappa l_i + \ti_i) - \kappa l_j (k^s \hcvd_s l_i) - \ti_j (k^s \hcvd_s l_i) \nonumber \\
& -l_j \Big(l^r \hcvd_r (k^s \hcvd_s l_i)\Big) - (\hw_j - \hp_j)q^r_{~i} \ti_r - k_j \Big(l^s \hcvd_s(q^r_{~i}\ti_r)\Big) ~.
\label{manip22}
\end{align}
Projection of the above Eq. \eqref{manip22} onto the $2$-surface $S_t$ leads to,
\begin{align}
q^i_{~m} q^j_{~n}l^k \hcvd_k (\hcvd_j l_i)  = q^i_{~m} q^j_{~n}(l^k \hcvd_k \hphi_{ij}) - (q^j_{~n}\ti_j)(q^i_{~m}k^s \hcvd_s l_i) +\hp_n (q^r_{~m}\ti_r) ~.
\label{manip23}
\end{align}
We equate the Eqs. \eqref{manip23} and \eqref{manip21} following \eqref{tidal1}. Upon using the fact that,
\begin{align}
q^i_{~m} q^j_{~n} \hcvd_j (q^r_{~i}\ti_r) = \hsvd_n (q^r_{~m}\ti_r) ~,
\end{align}
we end up after some simplification, with the following result,
\begin{align}
q^i_{~m}q^j_{~n} (l^k \hcvd_k \hphi_{~ij}) &= - \hphi_{mt} \hphi^t_{~n} + (\kappa - \kt) \hphi_{mn} - \qq \hr_{risj}l^r l^s - \hhw_n (q^r_{~m}\ti_r) \nonumber \\
& + \hsvd_{n}(q^r_{~m}\ti_r) - \hphi_{mr}T^r_{~~kj}l^k q^j_{~n} ~.
\label{covhphi}
\end{align}
Essentially, we would want to convert the covariant derivative of the spatial tensor $\hphi_{ab}$ along the null generator into its Lie derivative counterpart. To that effect, in the RC spacetime $\sptm$, it is quite easy to establish that for any spatial tensor, we have,
\begin{align}
\lil \hphi_{ij} = l^r \hcvd_r \hphi_{ij} + l^r \hphi_{kj}T^k_{~~ri} + l^r \hphi_{ik}T^k_{~~rj} + \hphi_{kj}(\hcvd_il^k) + \hphi_{ik}(\hcvd_j l^k) ~.
\label{lielspatial}
\end{align}
This leads to,
\begin{align}
\qq \lil \hphi_{ij} = &\qq (l^r \hcvd_r \hphi_{ij}) + \hphi_{kn} T^k_{~~ri}l^r q^i_{~m} + \hphi_{mk}T^k_{~~rj}l^r q^j_{~n}\nonumber \\
& + q^i_{~m} \hphi^k_{~n}(\hcvd_i l_k) + q^j_{~n} \hphi_m^{~~k}(\hcvd_j l_k) ~.
\label{manip24}
\end{align}
Again, using the relation for the covariant derivative of the null generator i.e. \eqref{relnnablaalbnchi} and the fact that $\hphi_{ij}$ is a completely transverse spatial tensor, we can easily show that,
\begin{align}
q^i_{~m} \hphi^k_{~n}(\hcvd_i l_k) + q^j_{~n} \hphi_m^{~~k}(\hcvd_j l_k) = \hphi_{im} \hphi^i_{~n} + \hphi_{mi} \hphi^i_{~n} ~.
\label{manip25}
\end{align}
Using \eqref{manip25} in \eqref{manip24}, we hence have,
\begin{align}
\qq (l^r \hcvd_r \hphi_{ij}) = \qq \lil \hphi_{ij} - \hphi_{kn}T^k_{~~ri}l^r q^i_{~m} - \hphi_{mk}T^k_{~~rj}l^r q^j_{~n} - \hphi_{im} \hphi^i_{~n} - \hphi_{mi} \hphi^i_{~n} ~.
\label{manip26}
\end{align}
Employing \eqref{manip26} in \eqref{covhphi} and simplifying a bit, we end up having, 
\begin{align}
\qq \lil \hphi_{ij} &= \hphi_{im} \hphi^i_{~n} + (\kappa - \kt) \hphi_{mn} - \qq \hr_{risj}l^r l^s\nonumber \\
& + \hphi_{kn}T^k_{~~ri}l^r q^i_{~m} - \hhw_n (q^r_{~m}\ti_r) + \hsvd_n (q^r_{~m}\ti_r) ~.
\label{lielhphi} 
\end{align}
The above Eq. \eqref{lielhphi} defines the dynamical (Lie) evolution (along the null generator $\bm{l}$ of $\hech$) of the spatial tensor $\hphi_{ij}$ as projected on to the two surface $S_t$. On taking the trace of \eqref{lielhphi}, it can perhaps be anticipated that it leads to the dynamical evolution equation of the trace of $\hphi_{ij}$ along the null generator. This we will proceed with and carry out explicitly. Let us then take the trace of the L.H.S of \eqref{lielhphi} i.e $q^{ij} \lil \hphi_{ij}$. Let us note that the trace of the spatial tensor $\hphi_{ij}$ is,
\begin{align}
g^{ij} \hphi_{ij} = q^{ij} \hphi_{ij} = q^{ij}(\hchi_{ij} - \tit_{ij}) = \thdl - q^{rs}K_{trs}l^t = \thdl - q^{rs}T_{rts}l^t ~.
\label{tracehphi}
\end{align}
Computing the trace of the L.H.S of \eqref{lielhphi}, we have,
\begin{align}
q^{ij} \lil \hphi_{ij} &= q^{ij} l^r \hcvd_r \hphi_{ij} + l^r \hphi^{ji}T_{jri} + l^r \hphi^{ji}T_{irj} + \hphi^{ij} (\hcvd_j l_i) + \hphi^{ji}(\hcvd_j l_i) \nonumber \\
& = l^r \hcvd_r (g^{ij}\hchi_{ij}) + l^r (\hchi^{ji} - \tit^{ji})(T_{jri} + T_{irj}) + (\hchi^{ij} - \tit^{ij})(\hchi_{ij} - \tit_{ij}) \nonumber \\
&~~ ~~(\hchi^{ji} - \tit^{ji})(\hchi_{ij} - \tit_{ij}) \nonumber \\
&= l^k \hcvd_k \Big(\thdl - q^{rs}K_{trs}l^t\Big) + 2 l^r \hchi^{ij}T_{irj} - l^r \tit^{ij} (T_{irj}+ T_{jri}) + 2 \hchi^{ij} \hchi_{ij} - \tit_{ij} (2 \hchi^{ij}) \nonumber \\
&~~~~ - (\tit_{ij}\hchi^{ij} + \tit_{ji}\hchi^{ji}) + \tit_{ij}(\tit^{ij}+ \tit^{ji}) ~.
\label{manip27}
\end{align} 
Upon using the irreducible decomposition of the deformation rate tensor $\hchi_{ij}$ i.e \eqref{irredecompdeform}, the above relation \eqref{manip27} can be expanded to,
\begin{align}
q^{ij} \lil \hphi_{ij} &= l^k \hcvd_k \Big(\thdl - q^{rs}K_{trs}l^t\Big) + (l^t q^{rs}T_{rts}) \thdl + 2 l^r (\shdf^{ij} T_{irj}) - l^r \tit^{ij}(T_{irj} + T_{jri}) \nonumber \\
& + (\thdl)^2 + 2 (\shdf_{ij})(\shdf^{ij}) -2 (q^{rs}K_{trs}l^t) \thdl - 4 (\shdf_{ij})\tit^{ij} + \tit_{ij}(\tit^{ij}+ \tit^{ji}) ~.
\label{traceLHS}
\end{align}
Let us then compute the trace of the terms present in the R.H.S of \eqref{lielhphi}.
We have then,
\begin{align}
q^{mn} \hphi_{im} \hphi^i_{~m} = \fr (\thdl)^2 + (\shdf_{ij})(\shdf^{ij}) - (q^{rs}K_{trs}l^t) \thdl - 2 (\shdf_{ij})\tit^{ij} + \tit_{ij} \tit^{ij} ~;
\label{manip28}
\end{align}
\begin{align}
q^{mn}(\kappa - \kt) \hphi_{mn} = (\kappa - \kt) \Big(\thdl - q^{rs}K_{trs}l^t\Big) ~;
\label{manip29}
\end{align}
\begin{align}
-q^{mn}q^i_{~m}q^j_{~n}\hr_{risj} l^r l^s = - \hr_{ij}l^i l^j ~;
\label{manip30}
\end{align}
\begin{align}
q^{mn} \hphi_{kn}T^k_{~~ri}l^r q^i_{~m} = \fr (l^t q^{rs}T_{rts}) \thdl + (\shdf^{ij})T_{irj}l^r - \tit^{ij}T_{irj}l^r ~;
\label{manip31}
\end{align}
\begin{align}
-q^{mn} \hhw_n (q^r_{~m}\ti_r) = - \hhw_n (q^{rn}\ti_r)~;
\label{manip32}
\end{align}
\begin{align}
q^{mn} \hsvd_n (q^r_{~m}\ti_r) = \hsvd_i (q^{ij}\ti_j) ~.\label{manip33}
\end{align}
We then add up \eqref{manip28}, \eqref{manip29}, \eqref{manip30}, \eqref{manip31}, \eqref{manip32} and \eqref{manip33} to obtain the trace of the R.H.S of \eqref{lielhphi}. Equating the resultant relation with \eqref{traceLHS}, we end up after some simplification,
\begin{align}
l^k \hcvd_k \thdl &= - \fr (\thdl)^2 - (\shdf_{ij})(\shdf^{ij}) + (\kappa - \kt) \Big(\thdl - q^{rs}T_{rts}l^t\Big) - \hr_{ij}l^i l^j \nonumber \\
& +l^k \hcvd_k (q^{rs} T_{rts}l^t) + \fr (q^{rs}T_{rts}l^t) \thdl - (\shdf^{ij})T_{irj}l^r + \tit^{ij}T_{jri}l^r + 2 (\shdf^{ij}) \tit_{ij} \nonumber \\
& - \tit^{ij} \tit_{ji} - q^{ij} \hhw_i \ti_j + \hsvd_j (q^{ij}\ti_i) ~.
\label{thdlevol}
\end{align}
The resulting equation provides the geometrical dynamics of the evolution of the expansion scalar $\thdl$ (corresponding to the outgoing null generators) along $\bm{l}$ and relates it to the quantity $\hr_{ij}l^i l^j$. This equation \eqref{thdlevol} can hence be identified as the null Raychaudhuri equation corresponding to a congruence of hypersurface orthogonal null generators $\bm{l}$. The above form of the NRE should be compared with Eq. A$18$ of \cite{Dey:2022qqp}. Let us mention a few structural differences between them. Here, the construction of the NRE has been done via the taking the irreducible decomposition of the deformation rate tensor $\hchi_{ij}$ i.e \eqref{irredecompdeform}. The deformation rate tensor by construction is a symmetric spatial tensor and hence $\shdf_{ij}$ represents its symmetric traceless part. As contrasted with Eq. A$18$ of \cite{Dey:2022qqp}, this form the NRE does not incorporate any antisymmetric rotation terms. Eq. A$18$ has been constructed with the dynamical variable $\hb_{ij}$ which is the projected deviation tensor. The irreducible decomposition of the the projected deviation tensor involves an antisymmetric traceless part $\whbl$ and hence in principle $\shdl \neq \shbl$. 

Let us get back to the dynamical equation involving the spatial tensor $\hphi_{ij}$ i.e. \eqref{lielhphi}. We focus on the term $\qq \lil \hphi_{ij}$ in the L.H.S of \eqref{lielhphi}. Using the definition of the deformation rate tensor $\hchi_{ij}$ (\eqref{deformationrate}) and its irreducible decomposition (\eqref{irredecompdeform}) its quite easy to show that,
\begin{align}
\qq \lil \hphi_{ij} = \fr q_{mn} \Big(\thdl\Big)^2 + \thdl (\shdf_{mn}) + \fr q_{mn}(l^r \hcvd_r \thdl) + \qq \lil \Big(\shdf_{ij}- \tit_{ij}\Big) ~. 
\label{liesigmalhs}
\end{align}
Now, we incorporate the form of the NRE given in \eqref{thdlevol} to put down the value of $l^r \hcvd_r \thdl$. This leads us to,
\begin{align}
\qq \lil \hphi_{ij} &= \qq \lil \Big(\shdf_{ij}- \tit_{ij}\Big) + \frac{1}{4} q_{mn} \Big(\thdl\Big)^2 + \thdl (\shdf_{mn}) - \fr q_{mn}\Big(\shdf_{ij}\shdf^{ij}\Big) \nonumber \\
&+ \fr q_{mn}(\kappa - \kt) \Big(\thdl - q^{rs}T_{rts}l^t\Big)  - \fr q_{mn} \hr_{ij}l^i l^j + \fr q_{mn} \Big(l^k \hcvd_k (\qtl)\Big) \nonumber \\
& + \frac{1}{4}q_{mn}\Big(\thdl (\qtl)\Big) - \fr q_{mn}\Big(\shdf^{ij}T_{irj}l^r\Big) + \fr q_{mn} \Big(\tit^{ij}T_{jri}l^r\Big) + q_{mn} (\shdf^{ij}\tit_{ij}) \nonumber \\
& - \fr q_{mn} (\tit^{ij}\tit_{ji}) - \fr q_{mn} \Big(q^{ij} \hhw_i \ti_j\Big) + \fr q_{mn} \Big(\hsvd_i(q^{ij}\ti_j)\Big) ~.
\label{liesigmalhs1}
\end{align}
Next, along the same lines as above, using \eqref{hphi} and \eqref{irredecompdeform}, we can expand the R.H.S of \eqref{lielhphi}. We have, as a result,
\begin{align}
&\hphi_{im} \hphi^i_{~n} + (\kappa - \kt) \hphi_{mn} - \qq \hr_{risj}l^r l^s + \hphi_{kn}T^k_{~~ri}l^r q^i_{~m} - \hhw_n (q^r_{~m}\ti_r) + \hsvd_n (q^r_{~m}\ti_r) \nonumber \\
& = \fr \thdl \Big(\qq T_{jri}l^r\Big) + \Big(q^i_{~m} \shdf^j_{~n}T_{jri}l^r\Big) - \Big(q^i_{~m} \tit^j_{~n}T_{jri}l^r\Big) + \frac{1}{4}q_{mn} \Big(\thdl\Big)^2 \nonumber \\
& + \thdl (\shdf_{mn}) - \fr \thdl (\tit_{mn} + \tit_{nm}) + \shdf_{im} \shdf^i_{~n} - \Big(\shdf_{im}\tit^i_{~n} + \shdf_{in}\tit^i_{~m}\Big) \nonumber \\
&+ \tit_{im} \tit^i_{~n} + \fr (\kappa - \kt) q_{mn} \thdl + (\kappa - \kt) \shdf_{mn} - (\kappa - \kt) \tit_{mn} \nonumber \\
& - \qq \hr_{risj}l^r l^s - \hhw_{n}q^r_{~m} \ti_r + \hsvd_n (q^r_{~m} \ti_r) ~.
\label{liesigmarhs1}
\end{align}
Now, all that we need to do is to invoke \eqref{lielhphi} and hence equate \eqref{liesigmalhs1} and \eqref{liesigmarhs1}. After the necessary simplification, we end up with,
\begin{align}
\qq \lil \Big(\shdf_{ij} - \tit_{ij}\Big) &= q_{mn} \Big(\shdf_{ij} \shdf^{ij}\Big) + (\kappa - \kt) \Big(\fr q_{mn}(\qtl) + \shdf_{mn} - \tit_{mn}\Big) \nonumber \\
& - \fr q_{mn} \Big(l^k \hcvd_k (\qtl)\Big) + \fr \thdl \Big(\qq T_{jri}l^r - \fr q_{mn}(\qtl) - (\tit_{mn}+ \tit_{nm})\Big) \nonumber \\
& + \Big(q^i_{~m} \shdf^j_{~n}T_{jri}l^r + \fr q_{mn} (\shdf^{ij}T_{jri}l^r)\Big) - \Big(q^i_{~m} \tit^j_{~n}T_{jri}l^r + \fr q_{mn} (\tit^{ij}T_{jri}l^r)\Big) \nonumber \\
& - \Big(\shdf_{im}\tit^i_{~n} + \shdf_{in}\tit^i_{~m} + q_{mn}(\shdf_{ij}\tit^{ij})\Big) + \Big(\tit_{im}\tit^i_{~n} + \fr q_{mn}(\tit^{ij}\tit_{ji})\Big) \nonumber \\
& - \Big(q^r_{~m} \hhw_n \ti_r - \fr q_{mn}(q^{ij}\hhw_i \ti_j)\Big) + \Big(\hsvd_n(q^r_{~m} \ti_r) - \fr q_{mn}\hsvd_i(q^{ij}\ti_j)\Big) \nonumber \\
& - q^b_{~m} q^d_{~n} \hr_{abcd}l^a l^c + \fr q_{mn} \hr_{ac}l^a l^c ~.
\label{lieshdf1}
\end{align} 
In the above, we have used the result, that for a spatial two dimensional symmetric tensor, one has, $\shdf_{im}\shdf^i_{~n} = \fr q_{mn}(\shdf_{ij}\shdf^{ij})$. Now, we would want to break the curvature tensor $\hr_{abcd}$ in $\sptm$ into the Riemannian part $R_{abcd}$ and the torsion part and similarly for the Ricci tensor $\hr_{ij}$. The decomposition of the Riemannian curvature tensor $R_{abcd}$ reads as,
\begin{align}
R^a_{~bcd} = C^a_{~bcd} + \fr \Big(R^a_{~c}g_{bd} - R^a_{~d}g_{bc} + R_{bd}\d^a_{~c} - R_{bc}\d^a_{~d}\Big) + \frac{1}{6}R \Big(g_{bc}\d^a_{~d} - g_{bd} \d^a_{~c}\Big) ~,
\label{curv3}
\end{align} 
where $C_{abcd}$ is the traceless part of $R_{abcd}$ called the Weyl tensor.
Employing \eqref{curv1}, \eqref{curv2} and \eqref{curv3}, its quite easy to verify that,
\begin{align}
&- q^b_{~m} q^d_{~n} \hr_{abcd}l^a l^c + \fr q_{mn} \hr_{ac}l^a l^c = -  q^b_{~m} q^d_{~n} C_{abcd}l^a l^c \nonumber \\
& - q^b_{~m} q^d_{~n}\Big((\hcvd_c K_{adb} - \hcvd_d K_{acb}) + T^i_{~~cd}K_{aib} + (K^i_{~~cb}K_{adi} - K^i_{~~db}K_{aci})\Big)l^a l^c \nonumber \\
& + \fr q_{mn}\Big(\hcvd_i K^i_{~~ca} + \hcvd_c T_a + T^i_{~~jc}K^j_{~~ia} + K^i_{~~ja}K^j_{~~ci}+ T_i K^i_{~~ca}\Big)l^a l^c ~.
\label{weyl}
\end{align}
Eq. \eqref{lieshdf1} coupled with Eq. \eqref{weyl} would then define the dynamical evolution of the shear tensor $\shdf_{ij}$ corresponding to the deformation rate tensor. Our final goal of arriving at the tidal equation involves converting the projected (onto the two-surface $S_t$) Lie derivative (along $\bm{l}$) of the quantity $(\shdf_{ij} - \tit_{ij})$ into the covariant directional derivative counterpart. To that end, after a few trivial manipulations involving \eqref{relnnablaalbnchi} and \eqref{irredecompdeform}, it can be shown that,
\begin{align}
&\qq \lil \Big(\shdf_{ij} - \tit_{ij}\Big) = \qq l^r \hcvd_r \Big(\shdf_{ij} - \tit_{ij}\Big) + q^i_{~m} \Big(\shdf^k_{~n} - \tit^k_{~n}\Big)T_{kri}l^r \nonumber \\
& + \Big(\shdf_m^{~~k} - \tit_m^{~~k}\Big)q_n^{~j}T_{krj}l^r + \thdl \Big(\shdf_{mn} - \tit_{mn}\Big) + 2 \shdf_{mk} \shdf^k_{~n} \nonumber \\
& - 2 \shdf_{mk}\tit^k_{~n} - \shdf^k_{~n}(\tit_{mk} + \tit_{km}) + (\tit_{mk} + \tit_{km}) \tit^k_{~n} ~.
\label{lietopar}
\end{align}
Finally, we equate \eqref{lietopar} with \eqref{lieshdf1} and use the relation \eqref{weyl}. After some simplification, we finally end up with \eqref{tidalforce}.
\bibliographystyle{elsarticle-num}
\bibliography{torsion_references_dns}
\end{document}